\documentclass[11pt]{article}
\usepackage[latin1]{inputenc}
\usepackage{amsmath}
\usepackage{amsfonts}
\usepackage{amssymb}
\usepackage{color}
\usepackage{epsfig}

\newlength{\bredde}
\def\slash#1{\settowidth{\bredde}{$#1$}\ifmmode\,\raisebox{.15ex}{/}
\hspace*{-\bredde} #1\else$\,\raisebox{.15ex}{/}\hspace*{-\bredde} #1$\fi}
\newcommand{\bmpa}{\bmp{3.5cm}\vspace{0.1cm}\be}
\newcommand{\empa}{\nonumber\ee\vspace{0.1cm}\emp}
\newcommand{\bmpb}{\bmp{2cm}\vspace{0.1cm}\be}
\newcommand{\empb}{\nonumber\ee\vspace{0.1cm}\emp}

\newcommand{\be}{\begin{equation}}
\newcommand{\ee}{\end{equation}}
\newcommand{\bea}{\begin{eqnarray}}
\newcommand{\eea}{\end{eqnarray}}
\newcommand{\nn}{\nonumber}
\newcommand{\la}{\lambda}
\newcommand{\bs}{\begin{split}}
\newcommand{\es}{\end{split}}
\newcommand{\nid}{\noindent}
\newcommand{\bmp}{\begin{minipage}}
\newcommand{\emp}{\end{minipage}}
\newcommand{\bmpe}{\bmp{6.5cm}\vspace{0.1cm}\be}
\newcommand{\empe}{\nonumber\ee\vspace{0.1cm}\emp}

\newcommand{\sect}[1]{\setcounter{equation}{0}\section{#1}}

\def\Tr{{\mbox{Tr}}}
\def\Pf{{\mbox{Pf}}}

\begin{document}
\topmargin -1.4cm
\oddsidemargin -0.8cm
\evensidemargin -0.8cm
\title{\Large\bf
Gap Probabilities in Non-Hermitian Random Matrix Theory}

\vspace{1.5cm}
\author{~\\{\sc G.~Akemann}, {\sc M.J.~Phillips}, and {\sc L.~Shifrin}
\\~\\
Department of Mathematical Sciences \& BURSt Research Centre\\
Brunel University West London\\
Uxbridge UB8 3PH, United Kingdom
}

\date{}
\maketitle
\vfill
\begin{abstract}
We compute the gap probability that a circle of
radius $r$ around the origin contains exactly $k$ complex eigenvalues.
Four different ensembles of random matrices are considered:
the Ginibre ensembles and their chiral complex counterparts, with both 
complex ($\beta=2$) or quaternion real ($\beta=4$) matrix elements.
For general non-Gaussian weights we give a Fredholm determinant or Pfaffian
representation respectively, depending on the non-Hermiticity
parameter.
At maximal non-Hermiticity, that is for rotationally invariant weights,
the product of Fredholm eigenvalues for $\beta=4$ follows from 
$\beta=2$ by skipping every second factor,
in contrast to the known relation for Hermitian ensembles.
On additionally choosing Gaussian weights we give new explicit expressions
for the Fredholm eigenvalues in the chiral case,
in terms of Bessel-$K$ and incomplete Bessel-$I$ functions. This compares to
known results for the Ginibre ensembles in terms of incomplete
exponentials.
Furthermore we present an asymptotic expansion of the logarithm of the gap
probability for large argument $r$ at large $N$ 
in all four ensembles, up to including
the third order linear term.
We can provide strict upper and lower
bounds and present numerical evidence for its conjectured values, depending on
the number of exact zero eigenvalues in the chiral ensembles.
For the Ginibre ensemble at $\beta=2$ exact results were previously derived
by Forrester.


\end{abstract}
\vfill

\thispagestyle{empty}
\newpage

\renewcommand{\thefootnote}{\arabic{footnote}}
\setcounter{footnote}{0}


\sect{Introduction}\label{intro}

Non-Hermitian Random Matrix Theory (RMT) as introduced by Ginibre \cite{Gin}
is almost as old as the classical Wigner-Dyson ensembles.
In the past decade it has enjoyed a revival of interest and many different
applications of it have been made, where we refer to \cite{FS} for
a recent review. However, most works have concentrated on computing the
spectral 
correlation functions,
culminating recently in the solution of the Ginibre ensembles of real
asymmetric 
matrices \cite{beta1list}.

In this article we will focus on the computation of gap probabilities from
which the distribution of individual eigenvalues or spacings
can be derived. Once
the spectral correlations of an ensemble are known in terms of a
kernel of (skew) orthogonal polynomials in the complex plane,
one can in principle express the gap probabilities in terms of a Fredholm
determinant involving this kernel, as we will show.
However, such expressions are in general not very explicit, unless the
Fredholm eigenvalues are known.
On the other hand an expansion of the Fredholm determinant
in terms of integrals over spectral
correlation functions converges very rapidly (see e.g. \cite{ABSW}),
but nevertheless
becomes cumbersome because of having 2D integrals in the complex plane.

Explicit results for the gap probability
were first derived for the $\beta=2$ Ginibre ensemble with unitary
invariance at maximal non-Hermiticity \cite{GHS}.  Here the
probability $E_0^{(2)}(r)$
that a circle of radius $r$ around the origin is empty is
given in terms of a product over incomplete exponentials.
The same quantity follows for the $\beta=4$ Ginibre class \cite{Mehta},
where every second term in the product is skipped.
This result was then used in \cite{GHS} for $\beta=2$
to compute the so-called level
spacing distribution in
the complex plane, by placing one eigenvalue at the origin and computing the
probability to find a second eigenvalue at radius $r$.
Due to the translational invariance with respect to the large-$N$ macroscopic
density being constant on a disc 
the corresponding repulsion of complex eigenvalues is supposed to hold
everywhere in the bulk \cite{GHS}.

The first aim of the present work is to extend the above results to the chiral
complex ensembles at $\beta=2$ and 4, computing
explicitly the product of Fredholm eigenvalues at finite-$N$. These ensembles
were introduced in
\cite{James} $(\beta=2)$ and \cite{A05} $(\beta=4)$. One motivation for this
is the application of non-Hermitian RMT to Quantum Chromodynamics
(QCD) with non-zero chemical potential, and we refer to \cite{A07mu} for
references and a review.
The RMT predictions in the complex plane can be compared to numerical
solutions of
QCD. The first comparison on $\mathbb{C}$ was done using the complex level
spacing distribution of \cite{GHS} of the $\beta=2$ Ginibre ensemble in the
bulk of the spectrum \cite{MPW}. However, in QCD chiral symmetry is very
important at the origin, and a successful comparison to individual complex
eigenvalues there was made very recently \cite{ABSW}. Our results for
the gap in the chiral complex $\beta=2$ class  valid at maximal
non-Hermiticity were already announced
there. We will give a derivation of
this result and extend it to $\beta=4$.

Apart from these explicit results we give a Fredholm
determinant (Pfaffian) form valid for a general weight function and
non-Hermiticity parameter at $\beta=2$ (4). 
This is shown to be equivalent to a (matrix)
eigenvalue equation involving the kernel of (skew) orthogonal polynomials.
For rotationally invariant weights it implies that 
the relation between the gap at $\beta=2$ and 4 skipping every second Fredholm
eigenvalue holds in general, extending this relation known 
for the Gaussian Ginibre case \cite{Mehta}.

The second part is then devoted to the asymptotic expansion of the
logarithm of $E_0^{(\beta)}(r)$
in all four Gaussian ensembles at maximal non-Hermiticity.
Our motivation here is to extend similar considerations done for Hermitian
RMT. These include the so-called Widom-Dyson constant to the third order, and
the first computation in \cite{dCM}
was made rigorous only very recently \cite{Igor}.

Our approach will be more heuristic in the sense that we give an exact
derivation of the first two terms using standard asymptotic techniques as the
Stirling formula. For our corresponding third order linear terms we can
provide strict upper and lower bounds for their coefficients, as well as
numerical evidence for their conjectured values. 
Our results are consistent with the exact results for the Ginibre ensemble at 
$\beta=2$ given in \cite{PF92} to the fourth order.
The chiral ensembles which are our main focus depend explicitly on the number
$\nu\frac{\beta}{2}$ of exact zero-eigenvalues
in the third order linear term.

Our paper is organised as follows. In Section \ref{Def} we
define our ensembles and correlation functions. Section
\ref{Results} first introduces Fredholm theory in 
the generic non-rotationally invariant case, followed by a
calculation of the new product formulae for $E_0^{(\beta)}(r)$
for both Gaussian chiral complex ensembles at maximal non-Hermiticity. 
Section \ref{asymp} is devoted to a detailed analysis of
our asymptotic expansion of $\log[E_0^{(\beta)}(r)]$
for Ginibre $\beta=2$ at large $N$, 
and Section \ref{other_ensembles} summarises and
compares the corresponding results 
for the other three ensembles, including a small radius expansion. 
Our concluding remarks in Section
\ref{conc} are followed by several appendices where some technical results are
collected.

\sect{Definitions}\label{Def}

In this section we collect together the definitions of the random matrix
ensembles to be considered,
starting with the Gaussian Ginibre ensembles and then moving to their
chiral complex
counterparts. We then define all density correlation functions, gap
probabilities and individual eigenvalue distributions and point out their
mutual relationships.

\subsection{Ginibre ensembles}

The Ginibre ensembles depending on a non-Hermiticity parameter $\tau$, also
known as Ginibre-Girko or elliptic ensembles, are defined as
(see e.g. \cite{Sommers88,FKS})
\be
{\cal Z}_{Gin}^{(\beta)}\ \sim\
 \int d\Phi
\exp\left[-\frac{N}{1-\tau^2}\Tr \Big(\Phi^\dag
  \Phi-\frac\tau2(\Phi^2+\Phi^{\dag\,2})\Big)\right]  \ ,\ \
\tau\in[0,1)\ .
\label{ZmatrixGin}
\ee
Here $\Phi$ is a complex non-Hermitian $N\times N$
matrix for $\beta=2$ or quaternion real
matrix for $\beta=4$.
This ensemble can also be thought of as a two-matrix model
$\Phi=H_1+ivH_2$, with $v=[(1-\tau)/(1+\tau)]^{\frac12}$, being composed of
two Hermitian (self dual) matrices $H_{1,2}$ with distribution
$\exp[-N\Tr H_j^2/(1+\tau)]$. In the limit $\tau\to1$ the Gaussian Unitary or
Symplectic Ensemble is recovered.

In both ensembles one can go to a complex eigenvalues basis
$z_{j=1,\ldots,N}$ of the matrix $\Phi$
\be
{\cal Z}_{Gin}^{(\beta)} \ \equiv\ \prod_{j=1}^N \int_{\mathbb{C}} d^2z_j\
w_{Gin}(z_j)\ {\cal J}_N^{(\beta)}(\{z\})
\label{ZevbGin}
\ee
with Jacobian
\bea
{\cal J}_N^{(\beta=2)}(\{z\}) &\equiv&
\prod_{k>l}^N |z_k-z_l|^2\ =\ |\Delta_N(\{z\})|^2\ ,
\label{Jb2}\\
{\cal J}_N^{(\beta=4)}(\{z\}) &\equiv&
\prod_{k>l}^N |z_k-z_l|^2\ |z_k-z_l^{\ast}|^2
\prod_{h=1}^N |z_h-z_h^{\ast}|^2\ =\  \prod_{h=1}^N (z_h-z_h^{\ast})
\Delta_{2N}(\{z,z^{*}\})\ .
\label{Jb4}
\eea
Both can be expressed through a Vandermonde determinant
$\Delta_N(z)$, but in a way different from the Hermitian ensembles with real
eigenvalues. In $\Delta_{2N}(\{z,z^{*}\})$ the variables are ordered as
$z_1,z_1^*,z_2,\ldots$. 
The exponential weight function is given simply by
\be
w_{Gin}(z)\
\equiv\ \exp\left[-\frac{N}{1-\tau^2}
\Big(|z|^2-\frac\tau2(z^2+z^{*\,2})\Big)\right]\ ,
\label{weightGin}
\ee
for both $\beta=2,4$. In the case $\tau=0$ called maximal non-Hermiticity
it becomes rotationally invariant, a case we will study in great detail later.

At $\beta=2$ the ensemble can be solved using Hermite polynomials as
orthogonal polynomials in the complex plane, and we refer to
\cite{FKS98} for an exhaustive discussion.
At $\beta=4$ the solution is given in terms of
skew orthogonal polynomials in the complex plane, again constructed from
Hermite polynomials, and a complete discussion is given in \cite{EKb4}.
At $\tau=0$ a basis of (skew) orthogonal polynomials is always constructed
from monic
powers instead of Hermite polynomials.

\subsection{Chiral ensembles}

Next we turn to the chiral complex counterparts of these Ginibre ensembles.
They are defined as Gaussian two-matrix models \cite{James,A05}
\be
{\cal Z}_{ch}^{(\beta)}\ \sim\
 \int d\Phi  d\Psi\
\exp\left[-\frac{N(1+\mu^2)}{4\mu^2}\Tr (\Phi^\dag \Phi\ +\ \Psi\Psi^\dag)
-\frac{N(1-\mu^2)}{4\mu^2}\Tr (\Psi\Phi\ +\ \Phi^\dag\Psi^\dag)
\right] \ ,\ \ \mu\in(0,1]\ .
\label{Zmatrixch}
\ee
Here $\Phi$ and $\Psi^\dag$ are two matrices of
rectangular size $(N+\nu)\times N$ with either complex $(\beta=2)$ or
quaternion real $(\beta=4)$ elements, without further symmetry properties.
The two matrices can be composed from two matrices $\Phi=iH_1+\mu H_2$,
and $\Psi=iH_1^\dag+\mu H_2^\dag$ with $H_{1,2}$
being non-Hermitian (quaternion real)
with distribution $\exp[-N\Tr H_j^\dag H_j ]$ each.
In the limit $\mu\to0$ leading to $\Phi=-\Psi^\dag$ we
recover the chiral Gaussian Unitary or Symplectic Ensemble \footnote{The
  non-Hermiticity parameters of the two sets of ensembles can be brought onto
  an equal footing by mapping $\mu^2=(1-\tau)/(1+\tau)$.}.

We are interested in the complex eigenvalues of the matrix
{\footnotesize$\Big( \begin{array}{cc}0 & \Phi \\
\Psi &0 \\\end{array}\Big)$}, 
or equivalently of $\Phi\Psi$. This change of variables
is tedious, by first going to complex eigenvalues of $\Phi$ and $\Psi$ each
and then integrating out one set. With details given in
\cite{James} ($\beta=2$) and \cite{A05} ($\beta=4$) we only give the answer
\be
{\cal Z}_{ch}^{(\beta)} \ \equiv\ \prod_{j=1}^N \int_{\mathbb{C}} d^2z_j\
w_\nu^{(\beta)}(z_j)\ {\cal J}_N^{(\beta)}(\{z^2\})
\label{Zevb}
\ee
with the following non-Gaussian weight function
\be
w_\nu^{(\beta)}(z)\
\equiv\ |z|^{\beta\nu+2}
K_{\frac12\beta\nu}\left(N|z|^2\frac{1+\mu^2}{2\mu^2}\right)
\exp\left[ N(z^2+z^{*\,2})\frac{1-\mu^2}{4\mu^2}\right]\ ,
\label{weight}
\ee
which now depends on $\beta$ and the number of exact zero-eigenvalues
$\nu\geq0$. We note that the Jacobian ${\cal J}$ only differs from the
Ginibre ensembles by inserting {\it squared} variables.

The chiral ensembles eq. (\ref{Zevb}) can be solved in terms of
Laguerre polynomials as orthogonal \cite{James,AOSV,A05} and skew orthogonal
\cite{A05} polynomials in the complex plane for $\beta=2$ and 4 respectively.

Regarding the weight we note that the
Bessel-$K$ function again becomes an exponential at
half integer values of the index, e.g.
$ K_{\pm\frac12}(x)=\sqrt{\frac{2\pi}{x}}\,e^{-x}\sim \lim_{x\to\infty}
K_\nu(x)$,
or in the asymptotic limit of large arguments which is also reached when
taking $\mu\to0$. In the opposite limit $\mu=1$ at maximal non-Hermiticity we
obtain again a rotationally invariant weight.

\subsection{Correlation functions}

Let us now turn to the correlation functions to be calculated, which we define
for arbitrary weight functions.
We define the $k$-th gap probability with respect to radial ordering to
give the probability that $k$ (independent) complex eigenvalues lie inside the
circle of radius $r$ around the origin, and $N-k$ lie outside.
Here we do not count those $\nu$ eigenvalues which are always zero by
construction in the chiral ensembles. For the chiral ensembles the definition
reads
\footnote{This definition of $E_k^{(\beta)}(r)$ differs from \cite{ADp}
by a factor $1/k!$.},
\be
E_k^{(\beta)}(r) \ \equiv\ \frac{1}{{\cal Z}^{(\beta)}_{ch}}\frac{N!}{(N-k)!k!}
\prod_{j=1}^k \int_0^r dr_j r_j
\prod_{l=k+1}^N \int_r^\infty dr_l r_l
\prod_{n=1}^N \int_0^{2\pi} d\theta_n w_\nu^{(\beta)}(z_n)
\ {\cal J}_N^{(\beta)}(\{z^2\})\ ,
\label{Ekbdef}
\ee
where we have switched to polar coordinates $z_n=r_n e^{i\theta_n}$. The
quantities for the
Ginibre ensembles are the same and simply obtained by replacing the last two
factors by $w_{Gin}(z_n){\cal J}_N^{(\beta)}(\{z\})$, the respective joint
probability distribution function (jpdf).

In eq. (\ref{Ekbdef}) we could have given a more general definition, by
choosing other sets (i.e. not necessarily concentric circles around the
origin) or by including 
an angular dependence. While this was briefly sketched in \cite{ABSW} we focus
here on quantities that we can compute most explicitly. We will come back to
this point in the next section.

From the $k$-th gap probability defined above, the radial distribution of
the $k$-th individual eigenvalue ordered with respect to radius follows 
recursively by a simple differentiation
\be
\frac{\partial}{\partial r}E_k^{(\beta)}(r) \ \equiv\ r\left(
p_k^{(\beta)}(r) -
p_{k+1}^{(\beta)}(r)\right)\ ,
\label{Ekpkrel}
\ee
setting $p_0^{(\beta)}(r)\equiv0$. In particular this implies for the first
eigenvalue that
$\frac{\partial}{\partial r}E_0^{(\beta)}(r)=-rp_{1}^{(\beta)}(r)$.
The relation eq. (\ref{Ekpkrel}) is easily inverted to
\be
p_k^{(\beta)}(r)\ =\ -
\frac{1}{r}
\frac{\partial}{\partial r}\sum_{l=0}^{k-1}E_l^{(\beta)}(r)\ .
\label{pkEkrel}
\ee
Obviously it should hold that the sum over all individual eigenvalues gives
the spectral density $R_1^{(\beta)}(z)$ to be defined below,
\be
\int_0^{2\pi}d\theta  R_1^{(\beta)}\left(z=re^{i\theta}\right)\ = \
\sum_{k=1}^Np_k^{(\beta)}(r)\ \ ,\ \ \mbox{with}\ \ \int_0^\infty dr\,r
p_k^{(\beta)}(r)=1 \ \ \forall k\ .
\label{R1pksum}
\ee
Here we have integrated over the angle as the $p_k^{(\beta)}(r)$ only depend
on the radius. They are normalised with the two-dimensional radial measure.
This relation will be used later to illustrate and check our
results for individual eigenvalues.
We now come to all the $k$-point density correlation functions defined as
\begin{equation}
  R_k^{(\beta)}(z_1,\ldots,z_k)
\ \equiv\ \frac{1}{{\mathcal Z}^{(\beta)}_{ch}} \frac{N!}{(N-k)!}
\prod_{j=k+1}^N  \int_{\mathbb{C}} d^2z_j
 \prod_{l=1}^N w_\nu^{(\beta)}(z_l)\ {\cal J}_N^{(\beta)}(\{z^2\}) \ ,
  \label{Rkbdef}
\end{equation}
with the same quantities for Ginibre defined using their respective jpdf.
In particular for $k=1$ the spectral density is normalised to the number
of eigenvalues: $\int_{\mathbb{C}} d^2z R_1^{(\beta)}(z)=N$.

All $k$-point density functions are known explicitly for finite $N$ for all
four ensembles defined above:
\be
R_k^{(\beta)}(z_1,\ldots,z_k)=
\mbox{(Q)det}_{1\leq i,j\leq k}[K_N^{(\beta)}(z_i,z_j^*)]\ ,
\label{Rksol}
\ee
where  $K_N^{(2)}(z_i,z_j^*)$ is the kernel of the corresponding
orthogonal polynomials in the complex plane for $\beta=2$.
In our Gaussian ensembles these are of Hermite \cite{FKS} type for Ginibre, or
Laguerre \cite{James,AOSV} for the chiral case.
For $\beta=4$ the quaternion determinant Qdet (or Pfaffian) is taken
over the $2\times2$ matrix kernel $K_N^{(4)}$
expressed in terms of the pre-kernel $\kappa_N(z_i,z_j)$ of the 
corresponding skew orthogonal polynomials in the
complex plane. For our examples these are
given again by Hermite \cite{EKb4} (Ginibre) or Laguerre \cite{A05} (chiral)
polynomials.

The $k$-th gap probability can be expressed in terms of
$n$-point density correlation functions as follows
\begin{equation}
  E_k^{(\beta)}(r)=\frac{1}{k!}\sum_{\ell=0}^{N-k}\frac{(-1)^\ell}{\ell!}
\prod_{j=1}^{k+\ell}\int_0^r dr_j r_j
\int_0^{2\pi} d\theta_j\
R_{k+\ell}(z_1,\ldots,z_{k+\ell})\ ,
  \label{Eexp}
\end{equation}
as was pointed out in \cite{ABSW} for the chiral complex ensembles at
$\beta=2$. The term in the sum with $k+\ell=0$ is set to unity.
In fact this expansion holds
for all 4 ensembles both at $\beta=2$ and 4, independent of the structure of
the jpdf (or even for real $\beta$ in case of Hermitian RMT).
This can easily be obtained using the same expansion as for real eigenvalues in
\cite{ADp}, inserting merely the definition (\ref{Ekbdef}), and
without using the solution eq. (\ref{Rksol}).
Moreover one can define a generating function by generalising the zero-th gap
probability to
\bea
E^{(\beta)}(r;\xi)&\equiv&
\frac{1}{{\cal Z}^{(\beta)}_{ch}}
\prod_{j=1}^N \left(\int_0^\infty dr_j r_j-\xi\int_0^r dr_j r_j\right)
\int_0^{2\pi} d\theta_j w_\nu^{(\beta)}(r_j e^{i\theta_j})
\ {\cal J}_N^{(\beta)}(\{z^2\})
\nn\\
&=&\sum_{\ell=0}^{N}\frac{(-\xi)^\ell}{\ell!}
\prod_{j=1}^{\ell}\int_0^r dr_j r_j
\int_0^{2\pi} d\theta_j\
R_{\ell}(z_1,\ldots,z_{\ell})\ ,
  \label{Egendef}
\eea
leading to
\be
E_k^{(\beta)}(r)= \frac{(-1)^k}{k!} 
\frac{\partial^k}{\partial \xi^k}E^{(\beta)}(r;\xi)
\Big|_{\xi=1}, \ \mbox{for}\ \ k=0,1,\ldots, N\ .
\label{Egenrel}
\ee
The importance of this relation will become evident when we give a product
representation in terms 
of the eigenvalues of a certain determinant.

Note that although eq. (\ref{Eexp}) is explicit with all terms on the right
hand side known, it contains multiple integrals of
determinants of an increasing size. As an approximation the sum can be
truncated after the first few terms, and it was found in \cite{ABSW} to
converge rapidly 
for $k=0$. However, even within this approximation
the few integrals to be done numerically become rapidly cumbersome.

\sect{Results for the gap probabilities}\label{Results}

In Subsections \ref{nonmax} and \ref{nonmaxb4} 
we give some general results for $E_0^{(\beta)}(r)$
valid for generic non-Hermiticity.
In the chiral case this includes a determinant or Pfaffian
formula for recursively known one-dimensional integrals.
Then in Subsections \ref{max} and \ref{maxb4}
we turn to maximal non-Hermiticity where the
determinants or Pfaffians can be diagonalised and an explicit product
representation is given for any matrix size.

\subsection{General case $\beta=2$}\label{nonmax}

We will start the discussion with $\beta=2$ for a general weight function in
the eigenvalues representation,
including the chiral and Ginibre case.
First the gap probability $E^{(2)}_0(r)$ is treated, and then through the
generating functional $E^{(2)}(r;\xi)$ all subsequent gaps.
Our derivation is in complete analogy to the real case, see Section 6.3
\cite{Mehta}, but we will repeat it here in a slightly modified version to
prepare for $\beta=4$ which is new. We have that
\be
E_0^{(2)}(r) \equiv \frac{1}{{\cal Z}^{(2)_{ch}}}
\prod_{l=1}^N \int_{{\mathbb C}\setminus {\cal C}_r}d^2z_l\
w_\nu^{(2)}(z_l)
\ {\cal J}_N^{(2)}(\{z^2\})
\ =\ \prod_{l=1}^N \int_{{\mathbb C}\setminus {\cal C}_r}d^2z_l
\left|\det_{1\leq j,k,\leq N}[\varphi_{j-1}(z_k)]\right|^2\ ,
\label{gap2step1}
\ee
where ${\cal C}_r$ denotes the circle of radius $r$ around the origin.
While this will become important in Subsections 
\ref{max} and \ref{maxb4}, throughout this
subsection we could choose the set to be general.
The Ginibre case is trivially obtained by choosing the other weight and
non-squared arguments inside the Jacobian.

Here we have also used that the normalising
partition function is given as follows in terms of the squared
norms $h_j$ of the
corresponding orthogonal polynomials $p_j(z)$, $\int
d^2z\,w_\nu^{(2)}(z)p_j(z)p_k(z^*)=h_j\delta_{jk}$, 
\be
{\cal Z}^{(2)}_{ch} \ =\ N! \prod_{j=0}^{N-1} h_j \ .
\label{Zb2result}
\ee
Examples for the orthonormalised wave functions 
$\varphi_j(z)=w_\nu^{(2)}(z)^{\frac12}h_j^{-\frac12}p_j(z)$ are
\bea
\varphi_{j\,ch}(z) &=& w_\nu^{(2)}(z)^{\frac12}h_j^{-\frac12}
L_j^\nu\left(\frac{Nz^2}{1-\mu^2}\right) \ ,
\label{wavech}\\
\varphi_{j\,Gin}(z) &=& w_{Gin}^{(2)}(z)^{\frac12}h_j^{-\frac12}
H_j\left(z\sqrt{\frac{N}{\tau}}\right) \ ,
\label{waveGin}
\eea
for the chiral ensemble and Ginibre ensemble, respectively.
Applying Gram's result, e.g. see Appendix A.12. of \cite{Mehta}, we obtain
\be
E_0^{(2)}(r)=\det_{1\leq j,k,\leq N}\left[ \delta_{jk}-
 \int_{{\cal C}_r}d^2z\ \varphi_{j-1}(z)\varphi_{k-1}(z^*)\right]=
\prod_{i=0}^{N-1}\left(1-\la_i^{(2)}\right)\ .
\label{pregap2}
\ee
This determinant has to be diagonalised.  
Although all integrals inside the matrix elements can be determined recursively
for the chiral Gaussian case, see Appendix \ref{Kintegrals}
for a slightly different framework,
this does not allow us to compute the eigenvalues
$\la_j^{(2)}$ explicitly. Alternatively these eigenvalues can
be obtained from the following integral equation
\bea
\la\psi(u)&=&  \int_{{\cal C}_r}d^2z\, K_N^{(2)}(u,z^*)\psi(z)\ ,
\label{beta2ev}\\
K_N^{(2)}(u,z^*)&\equiv& \sum_{j=0}^{N-1} \varphi_{j}(u)\varphi_{j}(z^*)\ .
\label{K2def}
\eea
This is seen as follows. Because the $\varphi_{j}(u)$ form a basis the
eigenfunction $\psi(u)$ can be expanded in terms of them. Here, we will
use a slightly more formal argument than in \cite{Mehta}, by using the
projection property of the kernel,
\be
\int_{\mathbb C}d^2z K_N^{(\beta)}(u,z^*)K_N^{(\beta)}(z,v^*)
=K_N^{(\beta)}(u,v^*)\ \ ,\ \beta=2,4\ \ .
\label{contractKb}
\ee
As indicated this property 
will generalise to $\beta=4$ (for the definition of the $\beta=4$
kernel see eq. (\ref{K4def})). Integrating $K_N^{(2)}(v,u^*)$ times
eq. (\ref{beta2ev}) with respect to $u$ 
and applying eq. (\ref{contractKb}) we have
\bea
\la \sum_{i=0}^{N-1} \varphi_{i}(v)\ c_i
&=&\int_{{\cal C}_r}d^2z\, K_N^{(2)}(v,z^*)\psi(z)=\la\psi(v)
\label{psiexp2}\\
\mbox{with}\ \ c_i&=& \int_{\mathbb C}d^2z\,
\varphi_{i}(u^*)\psi(u)\ ,
\label{cidef}
\eea
and thus the desired expansion of the eigenfunction (for $\la\neq0$).
Plugging this expansion back into eq. (\ref{beta2ev}) we have
\be
\la\sum_{i=0}^{N-1} c_i\varphi_{i}(u)=
\sum_{j=0}^{N-1} \varphi_{j}(u) \int_{{\cal C}_r}d^2z\  \varphi_{j}(z^*)
\sum_{i=0}^{N-1} c_i\varphi_{i}(z)\ .
\ee
The projection onto the coefficients $c_l$ can be achieved using the
orthonormality of the $\varphi_{i}(u)$, integrating both sides with
$\int_{\mathbb C}d^2u\, \varphi_{l}(u^*)$:
\be
\la c_l = \sum_{i=0}^{N-1}c_i \int_{{\cal C}_r}d^2z\  
\varphi_{l}(z^*)\varphi_{i}(z)
\ .
\ee
This equation has a solution if
\be
0=\det_{1\leq i,j\leq N}
\left[ \la\delta_{ij}-  \int_{{\cal C}_r}d^2z\  \varphi_{i-1}(z^*)
\varphi_{j-1}(z)
\right]\ .
\ee
The determinant of this Hermitian matrix is given by a polynomial of degree
$N$ in term of its real eigenvalues $\la_i^{(2)}$
\be
\det_{1\leq i,j\leq N}
\left[ \la\delta_{ij}-  \int_{{\cal C}_r}d^2z\  \varphi_{i-1}(z^*)
\varphi_{j-1}(z)
\right]\equiv \prod_{i=0}^{N-1}\left(\la-\la_i^{(2)}\right)\ .
\label{fredb2}
\ee
This determinant is called the Fredholm determinant, 
and its eigenvalues are the Fredholm
eigenvalues which obviously depend on $r$ here. Setting $\la=1$ we are back to
eq. (\ref{pregap2}) as we wanted to prove.
 
Moving to the generating functional we can repeat all the steps.
Multiplying all equations by $\xi$ and redefining $\la\to\la\xi$ it
trivially follows for the generating functional
\be
E^{(2)}(r;\xi) =\prod_{i=0}^{N-1}\left(1-\xi\la_i^{(2)}\right)\ .
\label{gap2prodxi}
\ee
Applying definition eq. (\ref{Egendef}) the following representation
of the $k$-th gap probability holds:
\bea
E_k^{(\beta)}(r) &=& \prod_{j=0}^{N-1} \left(1-\la_j^{(\beta)}\right)
\ \sum_{\{j_i\}}
\frac{\la_{j_1}^{(\beta)}}{1-\la_{j_1}^{(\beta)}}
\cdots\frac{\la_{j_k}^{(\beta)}}{1-\la_{j_k}^{(\beta)}} \ ,\ \
\beta=2,4\ ,
\label{Ekb}
\eea
where we have anticipated that the same result holds for $\beta=4$ (as will be
shown next).
The same result is equally true for Hermitian RMT, see \cite{Mehta}
where it was derived in a different way.
In eq. (\ref{Ekb})
the sum is over all possible permutations of subsets of $k$ out of $N$
indices, e.g. for $k=1$ it reads:
\be
E_1^{(\beta)}(r)\ = \ \prod_{j=0}^{N-1}
\left(1-\la_j^{(\beta)}\right)\sum_{\ell=0}^{N-1}
\frac{\la_{\ell}^{(\beta)}}{1-\la_{\ell}^{(\beta)}}\ .
\ee

\subsection{General case $\beta=4$}\label{nonmaxb4}

Let us turn to $\beta=4$. To the best of our knowledge the
relation between the Fredholm eigenvalues and an eigenvalue equation
involving the kernel is new. Looking at the Jacobian eq. (\ref{Jb4})
and replacing the
single Vandermonde by a determinant of monic, skew orthogonal 
polynomials and their complex
conjugates (see eq. (\ref{skewdef})) we can write for the gap probability 
\bea
E_0^{(4)}(r) &\equiv& \frac{1}{{\cal Z}^{(4)}_{ch}}
\prod_{l=1}^N \int_{{\mathbb C}\setminus {\cal C}_r}d^2z_l\
w_\nu^{(4)}(z_l)
\ {\cal J}_N^{(4)}(\{z^2\})\ ,
\nn\\
&=&
\frac{1}{{\cal Z}^{(4)}_{ch}}\prod_{l=1}^N
\int_{{\mathbb C}\setminus {\cal C}_r}d^2z_l (z_l^2-z_l^{*\,2})
w_\nu^{(4)}(z_l)
\det_{1\leq k\leq N,\,1\leq j\leq 2N,
}
\left[
\begin{array}{l}
q_{j-1}(z_k^2)\\ 
q_{j-1}(z_k^{2\,*})\\
\end{array}
\right].
\label{gap4step1}
\eea
The $2N$ rows alternate in the variables $z_k$ and $z_k^*$.
The $N$-fold integral can be reduced to a Pfaffian over single integrals using
the de Bruijn integration formula, see e.g in \cite{EKb4}, and we obtain
\be
E_0^{(4)}(r) = \frac{(2N)!}{{\cal Z}^{(4)}_{ch}}
\Pf_{1\leq k,l\leq 2N}\left[
\int_{{\mathbb C}\setminus {\cal C}_r}d^2z(z^{2}-z^{*\,2})w_\nu^{(4)}(z)
\Big( q_{k-1}(z)q_{l-1}(z^{*})- q_{k-1}(z^{*})q_{l-1}(z)\Big)
\right].
\label{pregap4}
\ee
This is the $\beta=4$ result corresponding to eq. (\ref{pregap2}), valid for
any weight function which we display explicitly here.
The normalising partition function given by the squared norms,
${\cal Z}^{(4)}_{ch}=(2N)!\prod_{j=0}^{N-1}h_j$, 
as can be seen from the same calculation but over the full complex plane,
using the
definition (\ref{skewdef}).
The Ginibre ensembles are trivially obtained by choosing
the corresponding weights and
non-squared arguments inside the Jacobian and in front of the weight.

Our next task is to relate the eigenvalues of the matrix inside the
Pfaffian to eigenvalues involving the $\beta=4$ matrix kernel
given by \cite{EKb4}
\be
K_N^{(4)}(z,u^*)\equiv (z^{*\,2}-z^2)^{\frac12}(u^{*\,2}-u^2)^{\frac12}
w_\nu^{(4)}(z)^{\frac12}w_\nu^{(4)}(u)^{\frac12}
\left(\begin{array}{cc}
\kappa_N(z^*,u) & - \kappa_N(z^*,u^*)\\
\kappa_N(z,u) & - \kappa_N(z,u^*)\\
\end{array}\right)\ ,
\label{K4def}
\ee
where the pre-kernel is defined as
\be
\kappa_N(z,u)\equiv \sum_{k=0}^{N-1}h_k^{-1}[q_{2k+1}(z)q_{2k}(u)
-q_{2k+1}(u)q_{2k}(z)] \ .
\label{kappa4def}
\ee
The polynomials we choose to be skew orthogonal with respect to the following
antisymmetric product:
\be
\int_{{\mathbb C}}d^2z(z^{*\,2}-z^2)w_\nu^{(4)}(z)
\Big[ q_{2k+1}(z)q_{2l}(z^{*})- q_{2k+1}(z^{*})q_{2l}(z)\Big]
=h_k\delta_{kl} \ ,
\label{skewdef}
\ee
and zero whenever two even or two odd polynomials are contracted.
The kernel eq. (\ref{K4def}) defined in this way then satisfies the
contraction property eq. (\ref{contractKb}) \cite{EKb4}.

We can now state the matrix eigenvalue equation that is needed to solve
eq. (\ref{pregap4}), 
\be
\la
\left(\begin{array}{cc}\psi(u^*)\\\psi(u)\end{array}\right)
=
\int_{{\cal C}_r}d^2z(z^{*\,2}-z^2)w_\nu^{(4)}(z)
\left(\begin{array}{cc}
\kappa_N(u^*,z) & - \kappa_N(u^*,z^*)\\
\kappa_N(u,z) & - \kappa_N(u,z^*)\\
\end{array}\right)
\left(\begin{array}{cc}\psi(z^*)\\\psi(z)\end{array}\right).
\label{beta4ev}
\ee
The two components are trivially related by complex conjugation and thus give
rise to the same $\la$. However, for the sequel it is useful to use this
matrix form. As for $\beta=2$ the generalisation to gaps with respect to other
sets is trivial.

To proceed the expansion of the eigenfunctions in analogy to
eq. (\ref{psiexp2}) can be achieved by multiplying eq. (\ref{beta4ev}) by
the matrix in eq. (\ref{K4def}) from the left, integrating over
$(u^{*\,2}-u^2)w_\nu^{(4)}(u)$
and using the contraction eq. (\ref{contractKb}) to arrive at
\be
\la \int_{{\mathbb C}}d^2u (u^{*\,2}-u^2)w_\nu^{(4)}(u)
\left(\begin{array}{cc}
\kappa_N(v^*,u)\psi(u^*)-\kappa_N(v^*,u^*)\psi(u)\\
\kappa_N(v,u)\psi(u^*)-\kappa_N(v,u^*)\psi(u)\\
\end{array}\right)=\ \la
\left(\begin{array}{cc}\psi(v^*)\\\psi(v)\end{array}\right) \ .
\ee
The desired expansion of the eigenfunction $\psi(v)$ in terms of the skew
orthogonal polynomials thus reads
\be
\psi(v)=\sum_{k=0}^{N-1}\frac{1}{h_k}[q_{2k+1}(v)c_{2k}-q_{2k}(v)c_{2k+1}]
\ee
with
\bea
c_{k} &=& \int_{{\mathbb C}}d^2u (u^{*\,2}-u^2)w_\nu^{(4)}(u)
[q_{k}(u)\psi(u^*)-q_{k}(u^{*})\psi(u)]\ .
\label{ckeo}
\eea
Since the even skew orthogonal polynomials $q_{2k}(v)$
are not uniquely determined (see \cite{EKb4}), 
then neither is this expansion. However, after
fixing the former there is no ambiguity in the definition of the $c_k$'s.

Next we can project onto the coefficients $c_l$. Multiplying
eq. (\ref{beta4ev}) with $(-q_{l}(u),q_{l}(u^*))$ and integrating over
$u$ with the corresponding measure the skew orthogonality 
eq. (\ref{skewdef}) leads to
\bea
\la c_{l} &=& \int_{{\cal C}_r}d^2z (z^{*\,2}-z^2)w_\nu^{(4)}(z)
\sum_{k=0}^{N-1}h_k^{-1}\Big[
[q_{2k+1}(z^{*})c_{2k}-q_{2k}(z^{*})c_{2k+1}]q_{l}(z)
\nn\\
&&-[q_{2k+1}(z)c_{2k}-q_{2k}(z)c_{2k+1}]q_{l}(z^{*})
\Big]\ .
\label{ck}
\eea
Splitting into odd and even this equation can be written in matrix form as
\bea
&&\la\left(\begin{array}{cc}c_{2l}\\-c_{2l+1}\end{array}\right)
\ =\ \int_{{\cal C}_r}d^2z
(z^{*\,2}-z^2)w_\nu^{(4)}(z)\sum_{k=0}^{N-1}h_k^{-1}
\label{}\\
&&\times\left(\begin{array}{cc}
q_{2k+1}(z^*)q_{2l}(z)  -q_{2k+1}(z)q_{2l}(z^*)
& q_{2k}(z^*)q_{2l}(z)  -q_{2k}(z)q_{2l}(z^*) \\
q_{2k+1}(z^*)q_{2l+1}(z)-q_{2k+1}(z)q_{2l+1}(z^*)
& q_{2k}(z^*)q_{2l+1}(z)-q_{2k}(z)q_{2l+1}(z^*) \\
\end{array}\right)
\left(\begin{array}{cc}c_{2k}\\-c_{2k+1}\end{array}\right).\nn
\eea
It has a solution if
the following determinant of an antisymmetric $2N\times2N$ matrix vanishes 
(after normalising the $q_k(z)$ with $h_k^{-\frac12}$):
\bea
0&=&\det_{l,k=0,N-1}\left[
\la\left(\begin{array}{cc}0&-\delta_{kl}\\
\delta_{kl}&0\\\end{array}\right)
-\int_{{\cal C}_r}d^2z (z^{*\,2}-z^2)w_\nu^{(4)}(z)\ h_k^{-\frac12}
h_l^{-\frac12}
\right.
\label{fredb4}\\
&&\left.\times\left(\begin{array}{cc}
 q_{2k}(z^*)q_{2l}(z)  -q_{2k}(z)q_{2l}(z^*)
&q_{2k+1}(z^*)q_{2l}(z)  -q_{2k+1}(z)q_{2l}(z^*)\\
q_{2k}(z^*)q_{2l+1}(z) - q_{2k}(z)q_{2l+1}(z^*)
&q_{2k+1}(z^*)q_{2l+1}(z)-q_{2k+1}(z)q_{2l+1}(z^*)\\
\end{array}\right)
\right]
\nn\\
&=&\prod_{j=0}^{N-1}\left(\la-\la_j^{(4)}\right)^2\ .
\label{Eb4square}
\eea
Here we have swapped the columns of the $2\times2$ matrices which is allowed
under the determinant.
Let us compare to eq. (\ref{pregap4}): using the skew orthogonality and taking
out the norms which cancels the normalising prefactor,
the matrix inside the Pfaffian there
agrees with the one in eq. (\ref{fredb4}), up to an overall irrelevant sign
as $\Pf(A)=\Pf(A^T=-A)$. Because of the double degeneracy of
the eigenvalues from eq. (\ref{Eb4square}) we obtain as a final result
for the gap probability and the generating functional 
\be
E_0^{(4)}(r)=\prod_{j=0}^{N-1}\left(1-\la_j^{(4)}\right)\ \ ,\ \
E^{(4)}(r;\xi) =\prod_{j=0}^{N-1}\left(1-\xi\la_j^{(4)}\right)\ .
\label{gap4prod}
\ee
We have checked that the 
same result can be derived independently as a quaternion determinant, 
using orthogonal quaternions instead.
It would be very interesting to relate the two gap probabilities
eq. (\ref{gap2prodxi}) 
for $\beta=2$ and eq. (\ref{gap4prod}) for $\beta=4$ for 
a general weight function and at general non-Hermiticity.
In the following subsection we will be able to find such a relation for
maximal non-Hermiticity by diagonalising the
determinant or Pfaffian. For the Gaussian weight we can then
explicitly compute the $\la_j^{(\beta)}$ as a
function of $r$.

\subsection{Maximal non-Hermiticity $\beta=2$}\label{max}

In this section we derive an exact product representation for the gap
probabilities for an arbitrary rotationally invariant weight function. The
corresponding one-dimensional integrals can be solved explicity for our
Bessel-$K$ weight function eq. (\ref{weight}) at $\mu=1$, 
and have been computed
previously for the Ginibre ensemble with 
exponential weight  eq. (\ref{weightGin})
in \cite{GHS,Mehta} at $\tau=0$.

We again start with $\beta=2$ and a general weight function, and consider the
Gaussian chiral and Ginibre weights as examples at the end.
For any rotationally invariant weight $w_{\nu}^{(2)}(|z|)$
(e.g. eq. (\ref{weight}) at $\mu=1$)
the orthogonal polynomials are given by monomial powers (we will use the chiral
case with squared variables here, the Ginibre class trivially follows).

This can be seen from the orthogonality relation
\be
\int_0^{2\pi}d\theta\ z^{2k}z^{*\,2l}\ =\ 2\pi\,
t^{4k}\delta_{kl}\ \ ,\ k,l=0,1,\ldots\ \ ,
\label{angleOP}
\ee
where $z=te^{i\theta}$. Thus the wave functions in
eq. (\ref{gap2step1}) only differ by their norm, defined as
\be
\int_{\mathbb{C}}d^2z w_{\nu}^{(2)}(|z|) z^{2k}z^{*\,2l}
=2\pi\, \delta_{kl} \int_0^\infty dt\, t\, t^{4k} w_{\nu}^{(2)}(t)
\equiv \delta_{kl} h_k
\label{norm}
\ee
with $\varphi_k(z)=h_k^{-\frac12}w_{\nu}^{(2)}(|z|)^{\frac12}z^{2k}$.

Consequently the determinant in eq. (\ref{pregap2}) becomes diagonal due to
eq. (\ref{angleOP}) even though we only integrate over a circle of radius
$r$.
Putting the two results together we can read off the Fredholm eigenvalues
for the gap probability
\be
E_0^{(2)}(r) \ =\  \prod_{j=0}^{N-1} \left(1-\la_j^{(2)}(r)\right)
\ ,\ \ \mbox{with}\ \
\la_j^{(2)}(r)\ \equiv\
\frac{\int_0^rdt\,t^{4j+1} w_\nu^{(2)}(t)}
{\int_0^\infty dt\,t^{4j+1} w_\nu^{(2)}(t)}\ .
\label{E0b2}
\ee

After this general result let us return to our two Gaussian matrix models as
examples. The weight function of the chiral model eq. (\ref{weight}) at
$\mu=1$ contains a Bessel-$K$ function and all
integrals can be done explicitly.
First we compute the norms using a standard
integral, see e.g. eq. (6.561.16)
\cite{Grad},
\be
\int_0^\infty ds\,s^{2k+\nu+1}
K_\nu(s) \ =\
2^{2k+\nu}(k+\nu)!k!\ ,
\label{normint}
\ee
after choosing the scaling variable $s=N|z|^2$.

The second integral needed is derived in Appendix \ref{Kintegrals}
\bea
F_\nu(k,x)\ \equiv\
\int_0^x ds s^{2k+\nu+1} K_\nu(s)&=&
2^{2k+\nu}(k+\nu)!k!\left(
1-\frac{x^{2k+\nu+1}}{2^{2k+\nu}(k+\nu)!k!}K_{\nu+1}(x)\right.\nn\\
&&\left.-\
x\Big(I_{\nu+2}^{[k-2]}(x)K_{\nu+1}(x)+I_{\nu+1}^{[k-1]}(x)K_{\nu+2}(x)\Big)_{}
\right).
\label{Kintb2}
\eea
Here we have introduced the incomplete Bessel-$I$ functions as
\be
I_\nu^{[k]}(x)\ \equiv\
\sum_{l=0}^k \frac{1}{(l+\nu)!l!}\left(\frac{x}{2}\right)^{2l+\nu}
\ ,\ \ \mbox{for}\ k\geq0 \ ,
\label{Iincompletedef}
\ee
and zero for a negative upper summation index. The normalisation
eq. (\ref{normint}) can be written as $F_\nu(k,x=\infty)$, and for the
gap probability we thus have
\be
E_0^{(2)}(r) \ =\
\prod_{k=0}^{N-1} \left(1-\frac{F_\nu(k,Nr^2)}{F_\nu(k,\infty)}\right).
\label{Eb2F}
\ee
Most explicitly we can thus write out one minus the Fredholm eigenvalues as
\be
1-\la_k^{(2)}(r)
=
\frac{x^{2k+\nu+1}}{2^{2k+\nu}(k+\nu)!k!}K_{\nu+1}(x)
+x\Big(I_{\nu+2}^{[k-2]}(x)K_{\nu+1}(x)
+I_{\nu+1}^{[k-1]}(x)K_{\nu+2}(x)\Big)\ ,
\label{frEig}
\ee
where $x=Nr^2$. Inserted into the generating functional
eq. (\ref{gap2prodxi}) this gives all
gap probabilities for our chiral ensemble with $\beta=2$.
This holds both at finite $N$ and infinite $N$, where the product extends to
infinity and the combination $x=Nr^2$ is kept fixed.
We also note that in the large-$k$ limit the second term in eq. (\ref{frEig})
relates to the
Wronsky identity for (complete) Bessel-$I$ functions,
\be
0\ =\ 1-x\Big(I_{\nu+1}(x)K_{\nu}(x)
+I_{\nu}(x)K_{\nu+1}(x)\Big)\ .
\label{wronsky}
\ee
For completeness we give the kernel at $\mu=1$ \cite{AOSV}
to be inserted in eq. (\ref{beta2ev}) 
the eigenvalues of which we have computed:
\be
K_N(z,v^*)=
|z|^{\nu+1}|v|^{\nu+1}K_{\nu}\Big(N|z|^2\Big)^{\frac12}
K_{\nu}\Big(N|v|^2\Big)^{\frac12}
\sum_{k=0}^{N-1}\frac{2N^{2k+2+\nu}}{\pi2^{2k+\nu}k!\,(k+\nu)!}z^{2k}v^{*\,2k}
\ . 
\label{kerexpl}
\ee

\begin{figure*}[htb]
\centerline{
\epsfig{file=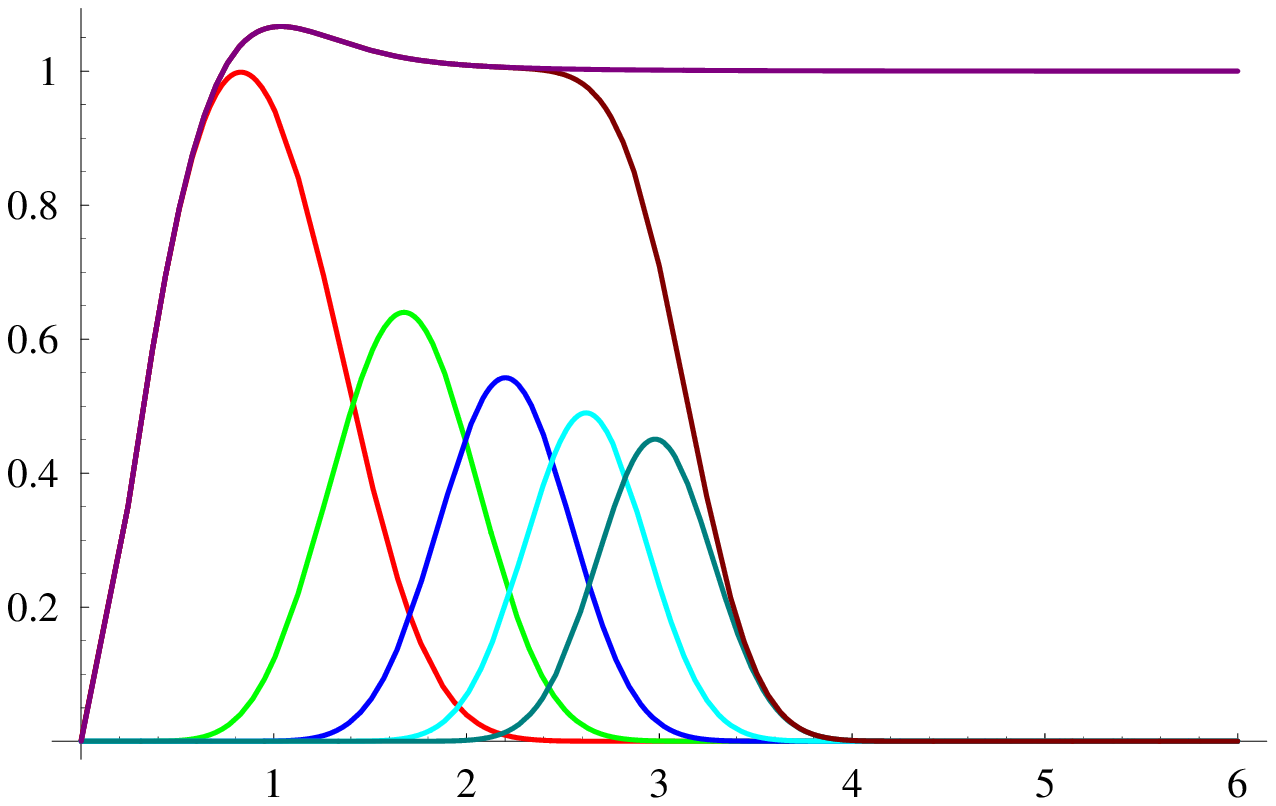,width=19pc}
  \epsfig{file=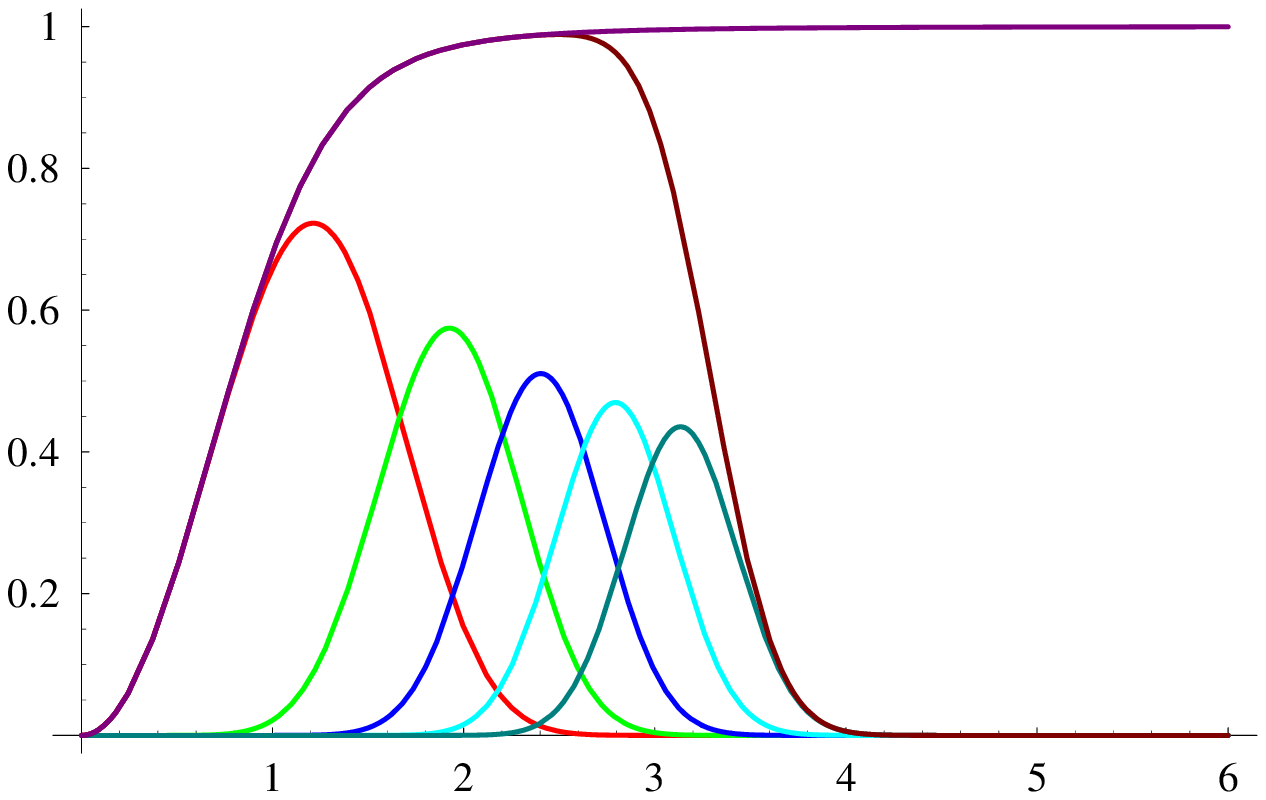,width=19pc}
\put(-210,160){$\pi\rho^{(2)}_1(s)$}
\put(-445,160){$\pi\rho^{(2)}_0(s)$}
\put(0,10){$s$}
\put(-230,10){$s$}
}
  \caption{The spectral density eq.~\eqref{rhob2}
of the chiral ensemble at $\beta=2$ times $\pi$, and the corresponding 
distributions of the
    first five eigenvalues eq.~\eqref{pkEkrel},
as well as their sum, for $\nu=0$ (left) and $\nu=1$ (right).}
  \label{psumb2}
\end{figure*}

Our second example is the known Ginibre ensemble at $\beta=2$ 
\cite{GHS}.
Instead of eq. (\ref{E0b2}) we have
\be
\la_{j\ Gin}^{(2)}(r)\ =\
\frac{\int_0^rdt\,t^{2j+1} w_{Gin}(t)}
{\int_0^\infty dt\,t^{2j+1}  w_{Gin}(t)}
\label{frEigGin}
\ee
from the weight eq. (\ref{weightGin}) at $\tau=0$. Consequently
each factor in the product for the gap probability can be written in terms of
(upper) incomplete Gamma functions:
\be
1-\lambda^{(2)}_{j\,Gin}(r)
\ =\
\frac{\Gamma(j+1,x)}{\Gamma(j+1)}
\ =\ e^{-x}\sum_{k=0}^{j}\frac{x^k}{k!} \ ,
\label{laginb2}
\ee
with $x=Nr^2$.
This result was first derived in \cite{GHS}, and the corresponding 
kernel in eq. (\ref{beta2ev}) 
reads \cite{FKS98}
\be
K_N(z,v^*)=\frac{N}{\pi}e^{-\frac{N}{2}(|z|^2+|v|^2)}\sum_{k=0}^{N-1}
\frac{N^k}{k!}z^kv^{*\,k}
\ .
\ee
Before turning to $\beta=4$ we would like to illustrate our new result
eq. (\ref{Eb2F}) 
by plotting the individual eigenvalues in the chiral case, using
eq. (\ref{R1pksum}).
For that purpose we compare the first few eigenvalues and their sum
with the large-$N$ microscopic spectral density at strong non-Hermiticity.
It is obtained from eq. (\ref{kerexpl}) \cite{AOSV,A03}
\be
\rho^{(2)}_\nu(s)\ \equiv\ \lim_{N\to\infty}\frac{1}{N}
R_1^{(2)}\left(|z|=\sqrt{\frac{s}{N}}\right)
\ =\ \frac{2}{\pi}s K_\nu(s)I_\nu(s)
\label{rhob2}
\ee
and has a very simple form depending only on the rescaled modulus $s$.
For large arguments it approaches the corresponding constant density of the
Ginibre ensemble, $\rho_{Gin}^{(2)}(s)=\frac{1}{\pi}$.
The corresponding
gap probabilities are now given by an infinite product. Because of its fast
convergence the individual eigenvalues can be computed by truncating
the products at $n$, where we have used $n=8$ in Figure \ref{psumb2}
to compute the first 5 eigenvalues from eqs. (\ref{Ekb}) and (\ref{pkEkrel}).
Note the normalisation eq. (\ref{R1pksum}) of the $p_k^{(2)}(r)$.
In Figure \ref{psumb2} we show $\pi$ times the density versus $\frac12
p_k^{(2)}(r)$.

\subsection{Maximal non-Hermiticity $\beta=4$}\label{maxb4}

Next we treat $\beta=4$ at maximal non-Hermiticity.
We choose the polynomials in eq. (\ref{gap4step1}) as monomial powers here
(these are not the skew orthogonal polynomials) as in the derivation
\cite{Mehta} 
\bea
\int_0^{2\pi}d\theta
(z^{*\,2}-z^{2})\Big( z^{2(k-1)}z^{*\,2(l-1)}- z^{2(l-1)}z^{*\,2(k-1)}\Big)
=4\pi \Big(t^{4k-4}\delta_{k-1,l}-t^{4k}\delta_{k,l-1}\Big),
\eea
where $z=te^{i\theta}$. Therefore the Pfaffian eq. (\ref{pregap4}) can be
computed as follows (see also \cite{Mehta} chapter 15), where every other term
contributes,
\be
E_0^{(4)}(r) \ =\ \prod_{j=0}^{N-1} \left(1-\la_{j}^{(4)}(r)\right)\ , \ \
\mbox{with}\ \ \la_{j}^{(4)}(r)\ \equiv\
\frac{\int_0^rdt\ t^{8j+5} w_\nu^{(4)}(t)}
{\int_0^\infty dt\ t^{8j+5} w_\nu^{(4)}(t)}\ .
\label{E0b4}
\ee
Here we have used the fact that the partition function ${\cal Z}^{(4)}_{ch}$
can be computed from eq. (\ref{pregap4}) at $r=0$.
This leads to the integrals in the denominator.
Eq. (\ref{E0b4}) is of the same form as eq. (\ref{E0b2}), with the difference
being that alternate powers are skipped. This leads to the relation
between the gap probabilities at $\beta=2$ and 4 true for general rotationally
invariant weight functions\footnote{We have to use the {\it same}
weight for both $\beta$ here, that is keeping $\beta\nu$ fixed in
eq. (\ref{weight}).}
\be
E_0^{(4)}(r)\ =\ \prod_{j=0}^{N-1}\left(1-\la_j^{(4)}\right)\ =\
\prod_{j=0}^{N-1}\left(1-\la_{2j+1}^{(2)}\right)
\label{E24rel}
\ee
where only the {\it odd} Fredholm eigenvalues from $\beta=2$ contribute to
$\beta=4$. The same statement can be made relating the generating
functionals, as well as for the Ginibre ensembles.
This new relation can be compared to the relation known for Hermitian
Gaussian RMT ($\tau=1$) in the large $N$ limit where one has
$E_0^{(4)}(r)=\frac12\Big(\prod_{i=0}^\infty(1-\la_{2i}^{(2)})+
\prod_{i=0}^\infty(1-\la_{2i+1}^{(2)})\Big)$, 
given as the {\it sum} of the product of even and odd
eigenvalues (see e.g. \cite{Mehta} eq. (11.7.5)).
It would be very interesting to generalise our relation eq. (\ref{E24rel}) to
the general case where one could take limits $\mu\to0$ or $\mu\to1$.
We can only conjecture that $\mu(\tau)$-dependent prefactors of a
general linear combination of the product of even and odd terms could
provide a form valid in the limiting cases.
For completeness we also give the expression for the Fredholm eigenvalues
in the Ginibre ensembles,
\be
\la_{j\ Gin}^{(4)}(r)\ =\
\frac{\int_0^rdt\,t^{4j+3} w_{Gin}(t)}
{\int_0^\infty dt\,t^{4j+3}  w_{Gin}(t)}\ .
\ee

Switching back to our explicit example of the chiral Gaussian model
with a Bessel-$K$ weight we do not need to do a new computation
and can simply use the integrals provided by eq. (\ref{frEig})
\be
E_0^{(4)}(r) \ =\ \prod_{k=0}^{N-1}
\left(1-\frac{F_{2\nu}(2k+1,Nr^2)}{F_{2\nu}(2k+1,\infty)}\right).
\label{E0prodb4}
\ee
In comparison to $\beta=2$ we have to shift the index $\nu\to2\nu$ here,
due to the explicit $\beta$-dependence of the weight eq. (\ref{weight}).
The corresponding pre-kernel to be inserted into eq. (\ref{beta4ev}) is given
by \cite{A05}
\be
\kappa_{N}(z,v^*)=
\frac{N^{2\nu+2}}{\pi2^{2\nu+3}}
\sum_{k=0}^{N-1}\sum_{j=0}^k
\frac{k!\Gamma(k+\nu+1)N^{2k+2j+1}}{\Gamma(2k+2\nu+2)(2k+1)!}
\frac{(z^{4k+2}v^{*\,4j}-z^{4j}v^{*\,4k+2})}{2^{4j}j!\Gamma(j+\nu+1)}
\ .
\label{kstrdef}
\ee

We also give the corresponding expression for the Ginibre
ensemble at $\beta=4$ which was already derived in \cite{Mehta},
\be
1-\lambda^{(4)}_{j\,Gin}(r)\ =\ 1-\la_{2j+1\,Gin}^{(2)}(r)\ =\
\frac{\Gamma(2j+2,x)}{\Gamma(2j+2)}
\ =\ e^{-x}\sum_{k=0}^{2j+1}\frac{x^k}{k!} \ ,
\label{laginb4}
\ee
with $x=Nr^2$. The pre-kernel is here given by \cite{EKb4}
\be
\kappa_{N}(z,v^*)=
\frac{N^{\frac32}}{2\pi}
\sum_{k=0}^{N-1}\sum_{l=0}^k \frac{N^{k+l+\frac12}}{(2k+1)!!(2l)!!}
(z^{2k+1}v^{*\,2l}-z^{2l}v^{*\,2k+1})\ .
\label{kstrdefGin}
\ee

\begin{figure*}[htb]
\centerline{
\epsfig{file=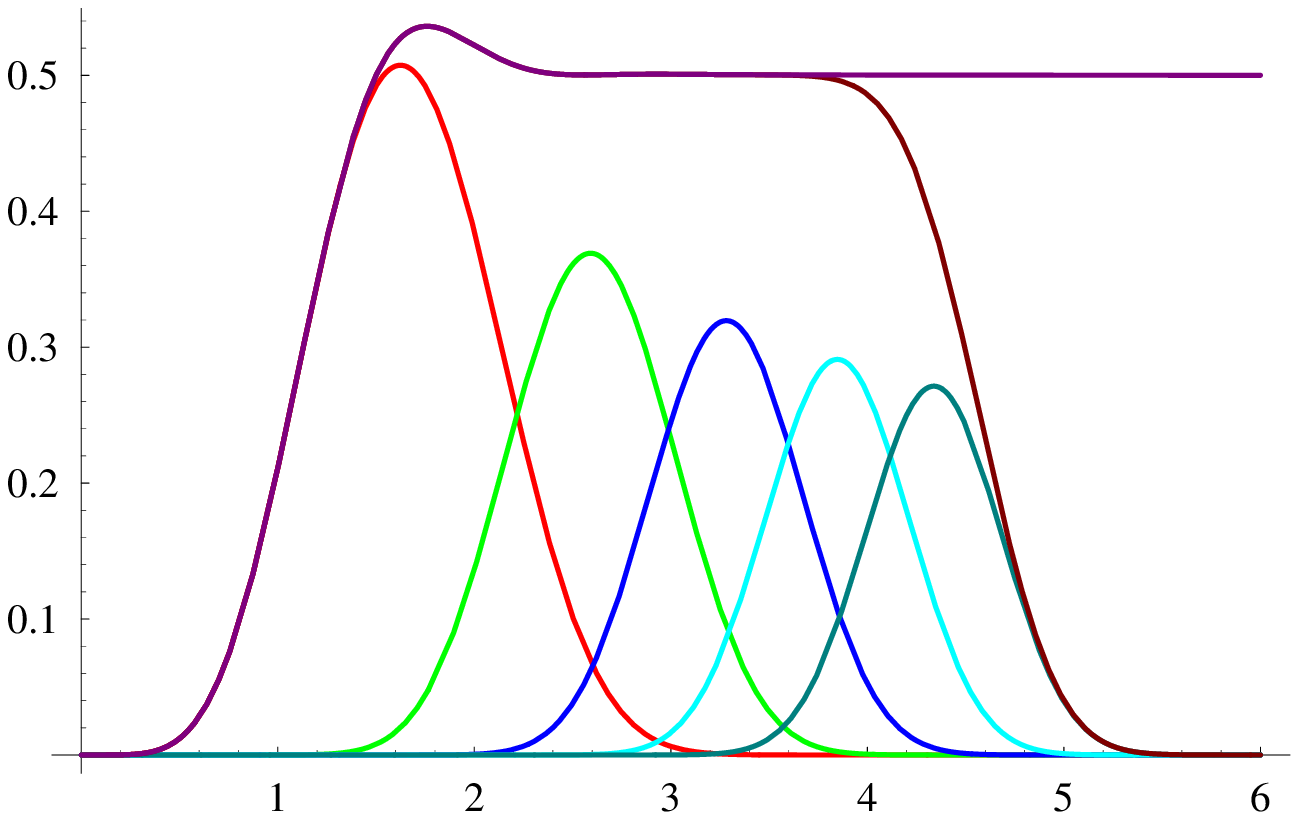,width=19pc}
  \epsfig{file=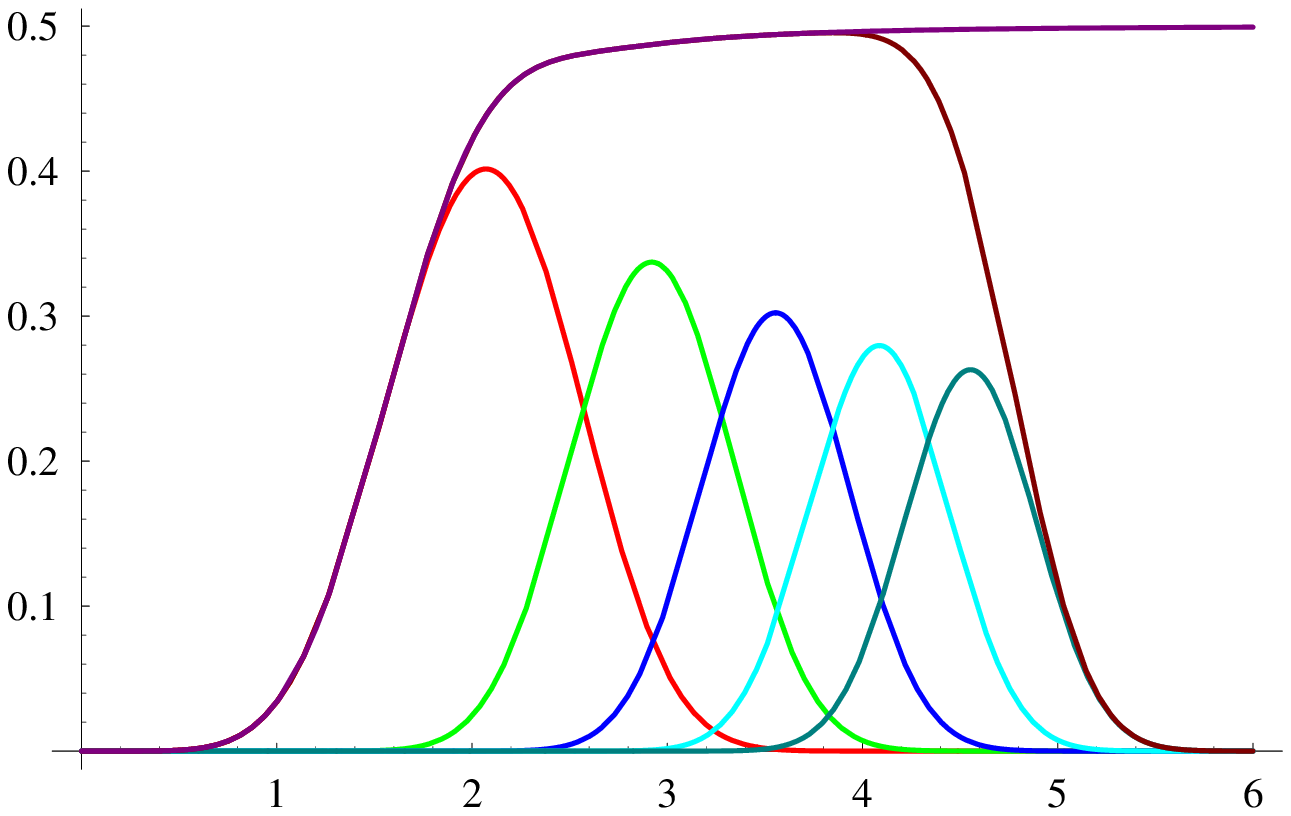,width=19pc}
\put(-230,150){$\int_0^{2\pi}\rho^{(4)}_1(se^{i\theta})d\theta$}
\put(-465,150){$\int_0^{2\pi}\rho^{(4)}_0(se^{i\theta})d\theta$}
\put(0,10){$s$}
\put(-230,10){$s$}
}
  \caption{The integrated spectral density eq.~\eqref{rhob4}
of the chiral ensemble at $\beta=4$, and distributions of the
    first five eigenvalues eq.~\eqref{pkEkrel},
as well as their sum, for $\nu=0$ (left) and $\nu=1$ (right).}
  \label{psumb4}
\end{figure*}

As before we illustrate the new individual eigenvalue distributions we found
for $\beta=4$ in the chiral case resulting from
eqs. (\ref{Ekb}) and (\ref{pkEkrel}).
The microscopic spectral density in the large-$N$ limit at maximal
non-Hermiticity is given by \cite{A05} 
(note the radius of support $\beta/2$ and height depend on $\beta$)
\be
\rho_{\nu}^{(4)}(\eta)
\ =\
\frac{1}{4\pi} (\eta^{\ast\,2}-\eta^2)|\eta|^2
K_{2\nu}\left({|\eta|^2}\right)
\int_0^1\frac{dt}{\sqrt{1-t^2}}\ I_{2\nu}(t|\eta|^2)
\sinh\left(\frac12\sqrt{1-t^2}\,(\eta^2-\eta^{\ast\,2})\right) ,
\label{rhob4}
\ee
where $\eta=\sqrt{N}r e^{i\theta}$. In contrast to $\beta=2$
the density is no longer rotationally
invariant and explicitly depends on the angle $\theta$. In particular it
vanishes along the real and imaginary axis, and for a more detailed discussion
of its symmetries we refer to \cite{A05}. In order to be able to compare with
the individual complex eigenvalues  we have to integrate over
the angle $\theta$ as indicated in eq. (\ref{R1pksum}). The result for the
first few eigenvalues is shown in Figure \ref{psumb4}.
Here we have again truncated the infinite product at $n=8$ to display the
first 5 eigenvalues and their sum.

\sect{Asymptotic expansion for the gap probability for Ginibre
  $\beta=2$}\label{asymp} 

We now turn to consider the large-$r$ asymptotics for the gap probability
$E_0^{(2)}(r)$  in the large $N$ limit keeping $x=Nr^2$ fixed.
Our method has the advantage that it can be easily extended to the chiral 
ensembles in the next section, although for this part it provides a weaker
result than given in \cite{PF92}.
Here we use eq. (\ref{E0b2}), but with the Ginibre
Fredholm eigenvalues 
from eq. (\ref{laginb2}).
Starting at unity, 
as $r$ increases $E_0^{(2)}(r)$ tends
towards zero.  Therefore, we define $P(x) > 0$ by
\be
E_0^{(2)}(r) \equiv \exp\left[-P(x=Nr^2)\right],
\ee
and will analyse the large-$x$ behaviour of $P(x)$, 
which is given by
\bea
\label{outer_sum}
P(x) & = & - \log \prod_{n=0}^{\infty} g_n(x) 
\ =\ - \sum_{n=0}^{\infty} \log g_n(x)
\eea
where
\be
\label{g_n_definition}
g_n(x) \equiv \sum_{k=0}^{n} t_k(x)
\ \ \mbox{and}
\ \ t_k(x) = e^{-x} \frac{x^k}{k!}.
\ee
The large $N$ limit of the corresponding Fredholm equation (\ref{beta2ev})
reads
\be
\la\psi(\eta)=\int_0^{x} dt\,t \int_0^{2\pi} d\theta\ 
e^{-\frac{1}{2}(t^2+|\eta|^2)+\eta te^{-i\theta}}
\psi(te^{i\theta})\ .
\label{b2FlargeN}
\ee


\subsection{Statement of large $x$ asymptotic results}

We show that using our method  

\vspace{0.3cm}
\noindent
\fbox{
\begin{minipage}{167mm}
\be
P(x)\ =\ \frac{x^2}{4} + \frac{x \log x}{2} + \left(\frac{\log
  2\pi}{2}-1\right)x + O\left(\sqrt{x}\right)
\ee
\end{minipage}
}\vspace{0.3cm}
although the coefficient of $x$ is partly conjectured.
It was derived rigorously in \cite{PF92} including the coefficient of the next
order term.
In more detail, we proceed by splitting the sum over $n$ in
eq. (\ref{outer_sum}) into two parts, namely $S_1({\cal N},x)$ (in which $n$
runs from 1 to ${\cal N}$) and $S_2({\cal N},x)$ 
(with $n$ running from ${\cal N}+1$ to $\infty$), where ${\cal N}$ is some
arbitrary (at 
this stage) parameter (which in general can depend on $x$)
\footnote{${\cal N}$ here is unrelated to the matrix size $N$ which has been
taken to infinity.}. 
We then further split $S_1({\cal N},x)$ into two parts, denoted $S_{11}({\cal
  N},x)$ and 
$S_{12}({\cal N},x)$, so that we have 
\bea
\label{P_definition}
P(x) & = & S_{11}({\cal N},x) + S_{12}({\cal N},x) + S_2({\cal N},x)
     \nonumber \\ 
     & \equiv & - \sum_{n=0}^{\cal N} \log t_n(x) - \sum_{n=0}^{\cal N} \log
     r_n(x) - 
     \sum_{n={\cal N}+1}^{\infty} \log g_n(x) 
\eea
\noindent with $r_n(x)$ defined as
\be
r_n(x) \equiv \frac{g_n(x)}{t_n(x)} =
1+\frac{n}{x}+\frac{n(n-1)}{x^2}+...+\frac{n!}{x^n}   \label{rndef} 
\ee
and $t_n(x)$ and $g_{n}(x)$ defined above in eq. 
(\ref{g_n_definition}). Since $S_{12}({\cal N},x)<0$, we
will 
always work with $|S_{12}({\cal N},x)|$ in the following.

With the choice that ${\cal N}=x$ (which we justify in Subsection \ref{split}),
we show 
that the following three properties {\it i) - iii)} hold:

\vspace{0.3cm}
\noindent
\fbox{
\begin{minipage}{167mm}
\bea
i) &&S_{11}(x,x) \ = \ \frac{x^2}{4} + \frac{x \log x}{2} + \frac{\log
  2\pi}{2}x + O(\log x)
\eea
\vspace{0cm}
\end{minipage}
}
\noindent
\fbox{
\begin{minipage}{167mm}
\bea
ii)&&S_{12}^{L\!B}(x,x) <  |S_{12}(x,x)|  < S_{12}^{U\!B}(x,x)
\eea
where
\bea
S_{12}^{U\!B}(x,x)  & = & x + O(\log x)
\eea
and
\bea
S_{12}^{L\!B}(x,x)  & = & \left[ 1 - \log 2 + \frac{\pi^2}{24} + \frac{1}{2}
  \textup{Li}_2(-e^{-2}) \right] x + O(1) 
\ \approx\ 0.653 x + O(1)
\eea
with $\textup{Li}_2(x)$ being the dilogarithm
\bea
\label{dilog_definition}
\textup{Li}_2(x) \equiv \sum_{k=1}^{\infty} \frac{x^k}{k^2} \ .
\eea
\vspace{0cm}
\end{minipage}
}
\noindent
\fbox{
\begin{minipage}{167mm}
\bea
iii) &&S_2(x,x) \ =\  O(\sqrt{x})\ .
\eea
\vspace{0cm}
\end{minipage}
}\vspace{0.3cm}
More precisely,
\be
0 < S_2(x,x) < S_2^{U\!B}(x)
\ee
where
\bea
S_2^{U\!B}(x) & = & M \sqrt{x} + O(1)
\eea
and $M$ is a constant given by
\bea
\label{M_def}
M & \equiv & - \int_0^{\infty} \log( \Phi(m)) dm 
\ \approx\ 0.478
\eea
and $\Phi(m)$ is the cumulative normal function
\be
\label{phi_definition}
\Phi(m) \equiv \frac{1}{\sqrt{2\pi}} \int_{-\infty}^m e^{-x^2/2} dx\ .
\ee

Furthermore, based on some preliminary numerical analysis, we
\textit{conjecture} the following somewhat stronger result for $S_{12}$ (we
comment on the bound for the sub-leading term later) 

\vspace{0.3cm}
\noindent
\fbox{
\begin{minipage}{167mm}
\bea
ii)'&& -S_{12}(x,x) \ = \ x +O(\sqrt{x})
\eea
\vspace{0cm}
\end{minipage}
}\vspace{0.3cm}
and for $S_2$ we conjecture

\vspace{0.3cm}
\nid
\fbox{
\begin{minipage}{167mm}
\bea
iii)' &&S_2(x,x) \ = \ M \sqrt{x} + C + o(1)
\eea
\vspace{0cm}
\end{minipage}
}\vspace{0.3cm}
where $M$ is defined as before (eq. (\ref{M_def})), and $C$ is another
constant, given by 
\bea
\label{C_def}
C & \equiv & \frac{1}{6\sqrt{2\pi}} \int_0^{\infty} \frac{(m^2-1)
   e^{-m^2/2}}{\Phi(m)} dm \ - \ \log 2 
\ \approx\ -0.716\ .
\eea


\subsection{Calculation of $S_{11}$}

We begin by computing $S_{11}$ containing the first two leading orders,
\begin{eqnarray}
S_{11}({\cal N},x)  & = & - \sum_{n=0}^{\cal N} \log \left( e^{-x}
             \frac{x^n}{n!} \right) 
\ =\ \sum_{n=0}^{\cal N} ( x - n \log x + \log n! )
             \nonumber        \\ 
             & = & ({\cal N}+1)x - \frac{{\cal N}({\cal N}+1)}{2} \log x +
             \sum_{n=1}^{\cal N} \log n! 
\end{eqnarray}
(since $\log 0! = \log 1 = 0$). We evaluate the sum of factorials as follows:
\begin{eqnarray}
\label{log_n_fac}
\sum_{n=1}^{\cal N} \log n!  & = & \sum_{n=1}^{\cal N} \sum_{k=1}^n \log k
\ =\ \sum_{k=1}^{\cal N} ({\cal N}-k+1) \log k
\ =\ ({\cal N}+1) \sum_{n=1}^{\cal N} \log n -
                      \sum_{n=1}^{\cal N} n \log n  \nonumber \\ 
                      & = & ({\cal N}+1) \log {\cal N}! - \sum_{n=1}^{\cal N}
                      n \log n. 
\end{eqnarray}
For the first term, we use Stirling's formula 
\be
\label{stirling}
\log {\cal N}! = \left({\cal N}+\frac{1}{2}\right) \log {\cal N} - {\cal N} +
\frac{\log 2\pi}{2} + O \left( \frac{1}{{\cal N}} \right), 
\ee
and for the second term, we apply the Euler-MacLaurin summation formula
(as hinted in \cite{GHS})
\begin{eqnarray}
\sum_{n=1}^{\cal N} f(n)  & = & \int_1^{\cal N} f(t) dt + \frac{f(1)+f({\cal
    N})}{2} + O(f'({\cal N})) 
\end{eqnarray}
which gives
\begin{eqnarray}
\label{sum_n_log_n}
\sum_{n=1}^{\cal N} n \log n & = & \int_1^{\cal N} t \log t dt + \frac{{\cal
    N} \log {\cal N}}{2} + O(\log 
{\cal N})    
\ =\ \left[ \frac{t^2 \log t}{2} -
                                 \frac{t^2}{4} \right]_1^{\cal N} +
    \frac{{\cal N} \log 
                                 {\cal N}}{2} + O(\log {\cal N})    \nonumber
    \\  
& = & \frac{{\cal N}^2 \log {\cal N}}{2} - \frac{{\cal N}^2}{4} +
                                 \frac{{\cal N} \log {\cal N}}{2} + O(\log
    {\cal N}).  
\end{eqnarray}
Inserting eqs. (\ref{stirling}) and (\ref{sum_n_log_n}) into
eq. (\ref{log_n_fac}) gives us 
\bea
\label{S11_N_x_result}
S_{11}({\cal N},x) &=& ({\cal N}+1)x - \frac{{\cal N}({\cal N}+1)}{2}\log x +
            \frac{{\cal N}^2\log {\cal N}}{2} \nonumber \\ 
            & & {} - \frac{3{\cal N}^2}{4} + {\cal N}\log {\cal N} + \left(
            \frac{\log 2\pi}{2} -1 
            \right){\cal N} + O(\log {\cal N}) 
\eea
and on setting ${\cal N}=x$, we arrive at
\begin{eqnarray}
S_{11}(x,x) & = & \frac{x^2}{4} + \frac{x \log x}{2} + \frac{\log 2\pi}{2}x +
O(\log x). 
\end{eqnarray}


\subsection{Calculation of $S_{12}$}

We turn now to the large $x$ behaviour of
\be
|S_{12}({\cal N},x)| \equiv \sum_{n=1}^{\cal N}\log r_n(x),
\ee
with $r_n(x)$ defined in eq. (\ref{rndef}).  We can write the sum starting at
$n=1$, because $\log r_0(x)=0$. 

Unfortunately, we have found no simple method for calculating $S_{12}$, and so
we determine some upper and lower bounds, and then conjecture the behaviour of
$S_{12}$ based on some numerical investigations. Note that, in our notation,
$S_{12}^{U\!B}$ ($S_{12}^{L\!B}$) is an upper (lower) bound for $|S_{12}|$
(rather than for $S_{12}$ itself), since $S_{12} < 0$. 


\subsubsection{Upper bound for $|S_{12}|$}

Let us first obtain an upper bound for $|S_{12}({\cal N},x)|$, assuming 
${\cal  N}<x$. We have 
\bea
r_n(x) &\equiv& 1 + \frac{n}{x} + \frac{n(n-1)}{x^2} + ... + \frac{n!}{x^n}
\nonumber \\ 
       &<& 1 + \frac{n}{x} + \left( \frac{n}{x} \right)^2 + ... + \left(
\frac{n}{x} \right)^n + ... 
\ =\ \frac{1}{1-\frac{n}{x}}
\eea
for $n<x$. Therefore
\bea
\log r_n(x) &<& - \log \left( 1 - \frac{n}{x} \right)
\eea
from which it follows that (for ${\cal N}<x$)
\be
|S_{12}({\cal N},x)| < S_{12}^{U\!B}({\cal N},x) \equiv -\sum_{n=1}^{{\cal N}}
\log \left( 1 - \frac{n}{x} \right). 
\ee
We approximate the sum using the Euler-MacLaurin summation formula. Since our
analysis is valid only for ${\cal N}<x$, we choose ${\cal N}=x-1$, and this
gives  
\bea
S_{12}^{U\!B}(x-1,x) &=& x - \frac{1}{2} \log x + O(1).
\eea
We consider the term $n={\cal N}(=x)$ separately; each term in the definition
of $r_n(x)$ is individually $\leq 1$, and there are $n+1$ such terms, so 
\be
\label{r_n_absolute_max}
0 < \log r_n(x) < \log(n+1) = O(\log x).
\ee
Hence we have
\bea
\label{S12_UB_Result}
S_{12}^{U\!B}(x,x) &=& x + O(\log x).
\eea

\subsubsection{Lower bound for $|S_{12}|$}

We have not been able to find a particularly tight lower bound for $|S_{12}|$
using elementary methods. Our best attempt involves the use of the following
lower bound for $r_n(x)$ (denoted $r_n^{L\!B}(x)$) which is strong for low
values of $n$ (indeed, we have equality for $n=1$ and 2), but rather weaker
when $n$ is closer to (but still less than) $x$:
\bea
r_n(x) &=&    1 + \frac{n}{x} + \frac{n(n-1)}{x^2} + ... + \frac{n!}{x^n}
\nonumber \\ 
       &\geq& \frac{1}{2} \left( 1 + \left( 1 + \frac{2}{x} \right)^n \right)
\ \equiv\ r_n^{L\!B}(x).
\eea
Therefore
\bea
\log r_n^{L\!B}(x) &=& \log \left( 1 + \left( 1 + \frac{2}{x} \right)^n
\right) - \log 2  
\ =\ \log \left( 1 + \frac{2}{x} \right)^n + \log \left( 1 +
\left( 1 + \frac{2}{x} \right)^{-n} \right) - \log 2  \nonumber \\ 
            &=& n \log \left( 1 + \frac{2}{x} \right) + \sum_{k=1}^{\infty}
\frac{(-)^{k+1}}{k} \left( 1 + \frac{2}{x} \right)^{-kn} - \log 2. 
\eea
Hence
\bea
\label{S12_LB_intermediate}
-S_{12}({\cal N},x) &>& S_{12}^{L\!B}({\cal N},x) 
\ \equiv\ \sum_{n=1}^{\cal N}\log r_n^{L\!B}(x)   \nonumber \\
             &=& \frac{{\cal N}({\cal N}+1)}{2} \log \left( 1 + \frac{2}{x}
             \right ) + 
             \sum_{n=1}^{\cal N} \sum_{k=1}^{\infty} \frac{(-)^{k+1}}{k}
             \left( 1 + 
             \frac{2}{x} \right)^{-kn} - {\cal N} \log 2. 
\eea
It is convenient at this stage to set ${\cal N}$ equal to $x$. Using the
             result that 
\be
\left(1+\frac{1}{a}\right)^a = e + O\left(\frac{1}{a}\right)
\ee
we can easily evaluate the first term of eq. (\ref{S12_LB_intermediate})
\bea
\frac{x(x+1)}{2} \log \left( 1 + \frac{2}{x} \right ) &=& (x+1) \log \left( 1
             + \frac{2}{x} \right )^{x/2}  
\ =\ (x+1) (1 + O(x^{-1}))  \nonumber \\
       &=& x + O(1).
\eea
The second term, which can be evaluated using the result (which we do not 
prove here)
\bea
\sum_{n=1}^{\cal N} \sum_{k=1}^{\infty} \frac{(-)^{k+1}}{k} \left( 1 +
             \frac{2}{{\cal N}} 
             \right)^{-kn} & = & \left[ \frac{\pi^2}{24} +
             \frac{1}{2}\textup{Li}_2(-e^{-2}) \right] {\cal N} + O(1)
\ \approx\ 0.346 {\cal N} + O(1),
\eea
is also linear, as indeed is the third term (which is simply $-x \log 2$).

Combining these results then gives
\be
S_{12}^{L\!B}(x,x) = \left[ 1 - \log 2 + \frac{\pi^2}{24} + \frac{1}{2}
  \textup{Li}_2(-e^{-2}) \right] x + O(1) 
\ee
where Li is the dilogarithm defined in eq. (\ref{dilog_definition}). The term
in square brackets is approximately equal to 0.653. 


\subsubsection{Conjecture for $|S_{12}|$ based on numerical analysis}
In the absence of a concrete analytical proof for the asymptotic limit
of $S_{12}$ itself, we undertook some elementary numerical analysis. In
particular, we considered
\be
A_{12}(x) \equiv \frac{|S_{12}(x,x)|}{x}
\ee
for various (necessarily finite) values of $x$. The function $A_{12}(x)$
begins at $\approx 0.69$, and is increasing.  It flattens very quickly,
and appears to be converging monotonically to 1. For example, we find
that $A_{12}(200\,000) \approx 0.996$.

Assuming that the coefficient of $x$ is indeed unity, we then
numerically investigated
\be
A_{12}^*(x) \equiv \frac{|S_{12}(x,x)| - x}{\sqrt{x}}.
\ee
for increasing $x$. It was not completely apparent that this converges
(we again went as far as $x=200\,000$), but it seems possible that it
does, implying that the sub-leading term is $O(\sqrt{x})$.

Putting these together, we anticipate that
\bea
|S_{12}(x,x)| & = & x + O(\sqrt{x}),
\eea
(i.e. that the true asymptote actually equals our earlier \textit{upper}
bound, at least to leading order). Considering the sub-leading term,
this might (at first sight) appear inconsistent with eq.
(\ref{S12_UB_Result}), since we have replaced $O(\log x)$ with
$O(\sqrt{x})$, but there is no contradiction. The more complete
statement of our conjecture is, in fact, that
\bea
|S_{12}(x,x)| & = & x - c \sqrt{x} + o(\sqrt{x})
\eea
where $c$ is some \textit{positive} constant.


\subsection{Calculation of $S_2$ using an upper bound}

We need only to provide an upper bound for $S_2(x)$, since our aim is merely
to show that $S_2(x)$ is smaller than linear in $x$, and we know that $S_2(x)
> 0$.


From eq. (\ref{P_definition}) we have
\be
S_2({\cal N},x) \equiv - \sum_{n={\cal N}+1}^{\infty} \log g_n(x).
\ee
For $n>x$, we have (see Appendix \ref{g_n_appendix} for details) a lower bound
for $g_n(x)$: 
\bea
g_n(x) &>& g^{L\!B}(m,x)  
\ \equiv\ \Phi(m) - \sum_{i=1}^{\infty} \frac{c_i(m)}{x^{i/2}} - \alpha
\sqrt{x}e^{-\beta x} - \frac{\gamma e^{-\delta x}}{ \sqrt{x}} 
\eea
where we have introduced a scaled variable
\be
m \equiv \frac{n-x}{\sqrt{x}}
\ee
and $\Phi(m)$ is the cumulative normal function. The $c_i(m)$ are
$m$-dependent numbers, and $\alpha$, $\beta$, $\gamma$ and $\delta$ are
constants. Therefore we have (on setting ${\cal N}=x$) 
\bea
\label{S2_result}
S_2(x,x)  &<&  S_2^{U\!B}(x)  \nonumber \\
          &\equiv& - \int_x^{\infty} \log(g^{L\!B}(m,x))\ dn   
\ =\ - \sqrt{x} \int_0^{\infty} \log(g^{L\!B}(m,x))\  dm
\eea
and on factorising out the $\Phi(m)$ and then expanding the logarithm, we
          arrive at 
\bea
S_2^{U\!B}(x)  &=& - \sqrt{x} \int_0^{\infty} \log ( \Phi(m))\ dm + O(1)  
\ \equiv\ M \sqrt{x} + O(1)
\eea
where
\bea
M & \equiv & - \int_0^{\infty} \log( \Phi(m))\ dm 
\ \approx\ 0.478\ ,
\eea
see also eq. (27) in \cite{PF92}.
For our purposes, it is sufficient that we have shown $S_2(x,x) =
O(\sqrt{x})$.  However, some more detailed analysis (supported by numerical
analysis) leads us to conjecture the following, much stronger, result: 
\bea
S_2(x,x) & = & M \sqrt{x} + C + o(1)
\eea
where
\bea
C & \equiv & \frac{1}{6\sqrt{2\pi}} \int_0^{\infty} \frac{(m^2-1)
  e^{-m^2/2}}{\Phi(m)} dm - \log 2 
\ \approx\ -0.716\ .
\eea


\subsection{Choice of the `split point' ${\cal N}$} \label{split}

Our final result must, of course, be independent of the choice of
${\cal N}$. However, if we had chosen ${\cal N}\ll x$, then we would 
not have been able to
bound $S_2$ in the way that we did, and it would not have been of smaller
order than $S_1$.  Conversely, if we had chosen ${\cal N}\gg x$, 
then our argument
for $S_{12}$ would not have been applicable, and this term would then have
been of higher order.  So, setting ${\cal N}=x$ is a pragmatic choice which
simplifies our analysis, giving $S_{11} \gg |S_{12}| \gg S_2$. 

We can understand this in a little more detail if we analyse what happens to
each of $S_{11}$, $S_{12}$ and $S_{2}$ as we increase the value of ${\cal N}$
by 1. For $S_{11}$, we can estimate the effect of this by partially
differentiating eq. (\ref{S11_N_x_result}) with respect to ${\cal N}$: 
\be
\frac{\partial S_{11}}{\partial {\cal N}} = {\cal N}\left( \log {\cal N} -
\log x \right) - \left( 
{\cal N}-x \right) + \log {\cal N} - \frac{1}{2}\log x + O(1). 
\ee
We can effectively `minimise' this by setting ${\cal N}=x$, which has the
effect of killing everything down to $O(\log x)$. So, by setting 
${\cal  N}=x+1$ 
instead of ${\cal N}=x$, we will only see an increase in $S_{11}$ of
$O(\log x)$, which we have already decided to call `small'. 

Of course, $S_{12}$ must decrease by $O(\log x)$ to compensate, and the
argument preceeding eq. (\ref{r_n_absolute_max}) shows that this is precisely
the case.  ($S_2$ will change only by $O(1)$, and so can be ignored.) 

So, with this choice of ${\cal N}$, we can be sure that all of the `large'
contributions are already included in $S_{11}$; consequently, it is the
optimal choice. 


\sect{Asymptotic expansion of $E_0^{(\beta)}(r)$ 
for the other ensembles}\label{other_ensembles}

We now extend the problem to the $\beta=4$ case, and to the corresponding
chiral ensembles. For completeness, we will also include here the $\beta=2$
Ginibre case which was discussed in fuller detail in the previous section. 



The aim here, as earlier, is to determine $P(x)$ where $E_0(r) =
\exp(-P(Nr^2))$. We established before that 
\bea
g_n^{Gin}(x)        & = & \sum_{k=0}^{n} e^{-x} \frac{x^k}{k!},
\label{g_n_Gin_def} \\ 
g_{\nu,n}^{ch}(x) & = & \frac{x^{2n+\nu+1}K_{\nu+1}(x)}{2^{2n+\nu}n!(n+\nu)!}
+ x \left[ K_{\nu+1}(x)I_{\nu+2}^{[n-2]}(x) + K_{\nu+2}(x)I_{\nu+1}^{[n-1]}(x)
  \right] \label{g_n_ch_def} 
\eea
where $I_{\nu}^{[n]}(x)$ is the incomplete $I$-Bessel function defined in
eq. (\ref{Iincompletedef}). 
\vspace{0.5cm}
\noindent
\begin{table}[t]
\begin{tabular}{c|c|c}
Ensemble & $\beta=2$ & $\beta=4$ \\
\hline
Ginibre & \bmpe P(x) = -\sum_{n=0}^{\infty} \log g_n^{Gin}(x) \empe & \bmpe
P(x) = -\sum_{n=1,\;n\;\textup{odd}}^{\infty} \log g_n^{Gin}(x) \empe\\ 
\hline
Chiral & \bmpe P(x) = -\sum_{n=0}^{\infty} \log g_{\nu,n}^{ch}(x) \empe &
\bmpe P(x) = -\sum_{n=1,\;n\;\textup{odd}}^{\infty} \log g_{2\nu,n}^{ch}(x)
\empe\\ 
\end{tabular}
\caption{Definition of the exponent $P(x)$ for all four
  ensembles.}\label{Pdeftab} 
\end{table}

Our task is to find the asymptotic behaviour of each of the $P(x)$ 
which have the form 
given in Table \ref{Pdeftab}.
For $\beta=2$ chiral the 
large $N$ limit of the Fredholm equation (\ref{beta2ev}) takes the compact form
\be
\la\psi(\eta)=\int_0^{x} dt\,t \int_0^{2\pi} d\theta\ 
t|\eta| K_\nu(t^2)^{\frac12}K_\nu(|\eta|^2)^{\frac12}
I_\nu(\eta te^{-i\theta})\psi(te^{i\theta})\ ,
\label{b2chFlargeN}
\ee
after inserting the radial Bessel kernel. For $\beta=4$ a similar equation can
be written, but the corresponding kernels from eqs. (\ref{kstrdef}) and 
(\ref{kstrdefGin}) do not simplify.


\subsection{Statement of results}

To proceed, we first identify the leading term (highest power of $x$) of each
$g_n(x)$ 
\bea
t_n^{Gin}(x)        & \equiv & e^{-x} \frac{x^n}{n!},  \\
t_{\nu,n}^{ch}(x) & \equiv & \sqrt{\pi} e^{-x}
\frac{(\frac{x}{2})^{2n+\nu+\frac{1}{2}}}{n!(n+\nu)!} \label{T_ch_def} 
\eea
and denote the `remainder' as
\bea
r_n^{Gin}(x)        & \equiv & \frac{g_n^{Gin}(x)}{t_n^{Gin}(x)} \left( =
1+\frac{n}{x}+\frac{n(n-1)}{x^2}+...+\frac{n!}{x^n} \right),  \\ 
r_{\nu,n}^{ch}(x) & \equiv & \frac{g_{\nu,n}^{ch}(x)}{t_{\nu,n}^{ch}(x)}.
\eea
We have used the asymptotic form for the $K$-Bessel function
\be\label{K_Bessel_asymptotic}
K_{\nu}(x) \rightarrow \sqrt{\frac{\pi}{2x}}\;e^{-x}
\ee
as $x \rightarrow \infty$ when determining $t_{\nu,n}^{ch}(x)$ in
eq. (\ref{T_ch_def}); however, the terms that we dropped in doing this are
incorporated into $r_{\nu,n}^{ch}(x)$, so no approximation has been made. For
notational convenience, we drop the `$Gin/ch$' labels in what follows, and
write everything generically. 

We can write $P(x)$ as the sum of three sums:
\bea
S_{11}({\cal N},x)  & \equiv & - \sum_{0 \le n \le {\cal N}} \log t_n(x),  \\
S_{12}({\cal N},x)  & \equiv & - \sum_{0 \le n \le {\cal N}} \log r_n(x),  \\
S_2({\cal N},x)     & \equiv & - \sum_{n > {\cal N}} \log g_n(x)
\eea
where ${\cal N}$ is a suitably chosen `split point', and $n$ is restricted to
the odd 
integers for the $\beta=4$ ensembles. To simplify the appearance of our
results, we introduce $A(x)$, defined as 
\be
\label{A_def}
A(x) \equiv \frac{x^2}{4} + \frac{x \log x}{2} + \frac{\log 2\pi}{2}x \ .
\ee
\begin{table}
\begin{tabular}{c|c|c|c|c}
Ensemble & Choice of ${\cal N}$ & $S_{11}({\cal N},x)$ & $S_{12}({\cal N},x)$
& Coeff. of $x$ in $P(x)$ \\ 
\hline
Ginibre ($\beta=2$) & \bmpb x \empb & \bmpa A(x) \empa  & \bmpb -x \empb &
\bmpa \frac{\log 2\pi}{2}-1 \empa  \\ 
Ginibre ($\beta=4$) & \bmpb x \empb & \bmpa \frac{A(x)}{2}-\frac{x}{4} \empa &
\bmpb -\frac{x}{2} \empb & \bmpa \frac{\log 2\pi}{4}-\frac{3}{4} \empa \\ 
\hline
Chiral ($\beta=2$)  & \bmpb \frac{x}{2} \empb & \bmpa
\frac{A(x)}{2}-\frac{\nu}{2}x \empa & \bmpb -\frac{x}{2} \empb & \bmpa
\frac{\log 2\pi}{4}-\frac{\nu+1}{2}\empa \\ 
Chiral ($\beta=4$)  & \bmpb \frac{x}{2} \empb & \bmpa \frac{A(x)}{4}-\left(
\frac{2\nu}{4} + \frac{1}{4} \right) x \empa & \bmpb -\frac{x}{4} \empb & \bmpa
\frac{\log 2\pi}{8}-\frac{2\nu+2}{4} \empa   \\ 
\end{tabular}
\caption{Explicit results for the exponent $P(x)$ for all four
  ensembles.}\label{Presults} 
\end{table}
Our results are then summarised in Table \ref{Presults}
where we omit terms of smaller order than linear in $x$. For
$S_{11}$ these terms are $O(\log x)$, and for $S_{12}$ they are
$O(\sqrt{x})$. In all cases, $S_2(x) = O(\sqrt{x})$. 


\subsection{Derivation by relating the ensembles}

The method for each of the 3 other ensembles is broadly similar to that for the
original (Ginibre $\beta=2$) case, and so we do not give details here.
Essentially, the same techniques can be used to determine the relevant bounds,
and we have undertaken numerical analysis to support the results that we claim
above. 

However, it is quite instructive to understand the relationships
\textit{between} the results for the different symmetry classes, so we will
provide here some alternative (and quite concise) proofs of the $S_{11}$'s
which highlight these relationships. 


\subsubsection{From Ginibre to chiral}

Assuming the Ginibre results, we will show how to derive the corresponding
chiral results 
(for $\nu=0$ only), starting with the $\beta=2$ case. 
We begin
by stating the following exact relationship which follows immediately from the
definitions: 
\be
\log t_{0,n}^{ch}(x) = 2 \log t_n^{Gin}\left(\frac{x}{2}\right) + \frac{1}{2}
\log \frac{x}{2} + \frac{1}{2} \log \pi. 
\ee
Therefore
\bea
S_{11}^{ch,\beta=2}(x)  & \equiv & - \sum_{0 \le n \le x/2} \log
t_{0,n}^{ch}(x)  \nonumber \\ 
                  & = & - 2 \sum_{0 \le n \le x/2} \log
t_n^{Gin}\left(\frac{x}{2}\right) - \frac{1}{2} \sum_{0 \le n \le x/2} \left(
\log \frac{x}{2} + \log \pi \right)  \nonumber \\ 
                  & \equiv & 2 S_{11}^{Gin,\beta=2} \left( \frac{x}{2} \right)
- \frac{1}{2} \sum_{0 \le n \le x/2} \left( \log \frac{x}{2} + \log \pi
\right) \nonumber \\ 
                  & = & 2A\left( \frac{x}{2}\right) - \frac{x}{4} \left( \log
\frac{x}{2} + \log \pi \right) + O(\log x) 
\eea
where $A(x)$ was defined in eq. (\ref{A_def}). But it easily follows from the
definition of $A(x)$ that 
\be
2A\left(\frac{x}{2}\right) = \frac{A(x)}{2} + \frac{x}{4} \left( \log
\frac{x}{2} + \log \pi \right) + O(\log x) 
\ee
where the scaling of the leading term is explained by the fact that $A(x)$ is
essentially quadratic in $x$. Hence we have 
\bea
S_{11}^{ch,\beta=2}(x)  & = & \frac{A(x)}{2} + O(\log x)  \nonumber \\
                  & = & \frac{S_{11}^{Gin,\beta=2}(x)}{2} + O(\log x),
\eea
i.e. the chiral sum is half the Ginibre sum. This is the result given in
Table \ref{Presults} (when $\nu=0$). 

A similar argument gives the corresponding result for $\beta=4$ (at $\nu=0$):
\bea
S_{11}^{ch,\beta=4}(x) & = & \frac{S_{11}^{Gin,\beta=4}(x)}{2} - \frac{x}{8} +
O(\log x). 
\eea
Note that here we pick up an extra term which is linear in $x$, so it is not
true in general that the chiral case is always half the Ginibre case (at
least, not when we consider the terms linear in $x$). 


\subsubsection{From $\beta=2$ to $\beta=4$}\label{beta_2_to_beta_4}

The $\beta=4$ case involves summing alternate (odd $n$) terms from the same
sequence used for $\beta=2$. We would therefore expect the total for $\beta=4$
to be approximately half that for $\beta=2$, since $-\log g_n(x)$ is a
`smooth' function of $n$ (for fixed $x$). However, since $-\log g_n(x)$ is
monotonic decreasing as a function of $n$, we would expect the sum of the odd
terms (starting at 1) to be slightly less than the sum of the even terms
(which start at zero), and indeed we do see this bias. 

Let us now quantify this; we will do this first (and in detail) for the
Ginibre ensembles.  We have 
\bea
S_{11}^{Gin,\beta=2}(x) & \equiv &  - \sum_{0 \le n \le x} \log  \left( e^{-x}
\frac{x^n}{n!} \right)  \nonumber \\ 
                         & = & - \sum_{\substack{1 \le n \le x \\
    n\;\textup{odd}}} \left\{ \left( \log  e^{-x} \frac{x^{n-1}}{(n-1)!}
\right) + \log \left( e^{-x} \frac{x^n}{n!} \right) \right\}  \left({}+ O(\log
x) \;\textup{if}\;x\;\textup{even} \right) \nonumber \\  
                    & = & - \sum_{\substack{1 \le n \le x \\ n\;\textup{odd}}}
\left\{ 2 \left( \log  e^{-x} \frac{x^n}{n!} \right) - \log\left(  \frac{x}{n}
\right) \right\} \quad \left( {} + \;\textup{ditto} \right) 
\eea
and thus
\bea
S_{11}^{Gin,\beta=2}(x) & \equiv & 2 S_{11}^{Gin,\beta=4}(x) 
+ \sum_{\substack{1
    \le n \le x \\ n\;\textup{odd}}} \log \frac{x}{n} \quad \left( {} +
\;\textup{ditto} \right) \nonumber \\ 
                    & = & 2 S_{11}^{Gin,\beta=4}(x) + \sum_{0 \le n \le x/2}
\log x - \sum_{0 \le n \le x/2} \log (2n) + O(\log x)  \nonumber \\ 
                    & = & 2 S_{11}^{Gin,\beta=4}(x) + \frac{x}{2} + O(\log x)
\eea
where the last summation has been approximated using the Euler-MacLaurin
summation formula. A trivial rearrangement then gives\footnote{We could have
  changed the proportionality factor from $\frac12\to1$ by modifying
 the weight to $\exp[-\frac\beta2 N\Tr\ldots]$.} 
\be
S_{11}^{Gin,\beta=4}(x) = \frac{S_{11}^{Gin,\beta=2}(x)}{2} - \frac{x}{4} +
O(\log x) 
\ee
which is the result given in Table \ref{Presults}.

An almost identical argument gives a similar result for the chiral ensembles
(for $\nu=0$)$^5$: 
\be
S_{11}^{ch,\beta=4}(x) = \frac{S_{11}^{ch,\beta=2}(x)}{2} - \frac{x}{4} +
O(\log x)\ .
\ee


\subsection{Illustration of our results}

Writing eq. (\ref{g_n_Gin_def}) in terms of gamma functions (see
eq. (\ref{laginb2})) allows us to generalise $g_{n}^{Gin}(x)$ to non-integer
$n$. 
For the case when $x$ is large, we can then easily show that 
\be
g_{\nu,n}^{ch}(x) \approx g_{2n+\nu+\frac{1}{2}}^{Gin}(x)\ .
\label{shift}
\ee
The $P(x)$ for all of the four ensembles can therefore be written as products
of the logarithms of certain $g_n^{Gin}(x)$ as given in Table 
\ref{fredpicture}.

We can easily see from this why the Ginibre $\beta=4$ case is slightly less
than half of the Ginibre $\beta=2$ case, for example. There are half as many
dots in the second row as in the first, but they are systematically shifted to
the right (this is the same argument we presented above in Section
\ref{beta_2_to_beta_4}). 

More interesting is the comparison of the chiral and Ginibre cases for fixed
$\beta$. For the chiral $\beta=2$ case (for $\nu=0$, i.e. no exact zero
eigenvalues) there are also only half as many
dots, but there is no systematic shift compared with the Ginibre $\beta=2$
case as the chiral dots take the average value, e.g. $\frac12=(0+1/2)$ etc, 
and so the chiral  $S_1(x)$ is exactly half that of the Ginibre case. 
\vspace{0.5cm}
\noindent
\begin{table}[h]
\begin{tabular}{r|ccccccccccccccccc}
Ensemble & 0 & $\frac{1}{2}$ & 1 & $\frac{3}{2}$ & 2 & $\frac{5}{2}$ & 3 &
$\frac{7}{2}$ & 4 & $\frac{9}{2}$ & 5 & $\frac{11}{2}$ & 6 & $\frac{13}{2}$ &
7 & \textit{etc.} \\ 
\hline
Ginibre $\beta=2$  \hspace{1.0cm}     & $\bullet$ && $\bullet$ && $\bullet$ &&
$\bullet$ 
&& $\bullet$ && $\bullet$ && $\bullet$ && $\bullet$ \\ 
\hline
Ginibre $\beta=4$  \hspace{1.0cm}     &&& $\bullet$ &&&& $\bullet$ &&&&
$\bullet$ &&&& $\bullet$ \\ 
\hline 
\hline
Chiral $\beta=2$ ($\nu=0$)  && $\bullet$ &&&& $\bullet$ &&&& $\bullet$ &&&&
$\bullet$\\ 
($\nu=1$)  &&&& $\bullet$ &&&& $\bullet$ &&&& $\bullet$ \\
($\nu=2$)  &&&&&& $\bullet$ &&&& $\bullet$ &&&& $\bullet$ \\
\hline
Chiral $\beta=4$ ($\nu=0$)  &&&&&& $\bullet$ &&&&&&&& $\bullet$  \\
\hline
\end{tabular}
\caption{Pictorial illustration of the Fredholm eigenvalues taken by the four
  ensembles.} 
\label{fredpicture}
\end{table}

What about the case where we know there is exactly 
\textit{one} eigenvalue at the
origin? For the Ginibre $\beta=2$ case, we must then consider the probability
that, \textit{given there is one eigenvalue at the origin}, there are no other
eigenvalues within a distance $r$ of the origin. It turns out that this
conditional probability is calculated simply by removing the first ($n=0$)
term from the sum that defines $S_1$ (see also \cite{GHS}). 
Compare this modification of the top row
of Table \ref{fredpicture} 
with the chiral $\beta=2$ case for $\nu=1$ (i.e. when there is
again precisely one exactly-zero eigenvalue). It will be seen that there are
half as many dots in the latter case as in the former, and there is again no
systematic shift.
So, for precisely one exact zero eigenvalue, it is also
the case that the chiral case is exactly half the Ginibre case. 

It is not possible to do a similar comparison for $\beta=4$ directly, 
since the Ginibre
ensemble has zero probability to find an eigenvalue at the origin.


\subsection{Low $x$ asymptotics for the four
  ensembles}\label{low_x_asymptotics} 

The probability that one eigenvalue is at the origin and a second one at
radial distance $r$ can be used to compute the spacing distribution $p(s)$ in
the complex plane \cite{GHS}. Following their argument 
the mean of the macroscopic large-$N$
density is 
constant on a disc for the Ginibre ensemble $\beta=2$ 
and one can assume that the
probability calculated in this way is translation invariant, giving the
spacing everywhere in the bulk. 

However, for all the other ensembles this is not true and the origin is
special. What can be compared is the strength of repulsion between two complex
eigenvalues to leading order: for the Ginibre ensemble $\beta=2$ everywhere in
the bulk following \cite{GHS}, and for 
the chiral ensembles at the origin by placing an exact zero eigenvalue at zero.

\begin{table}[b]
\begin{tabular}{r|c|c}
Ensemble 
& $\beta=2$&  $\beta=4$  \\
\hline
Ginibre  \hspace{1.0cm}   &  \bmpe 1 - x + \frac{x^3}{2} -
\frac{5x^4}{12} + \frac{7x^5}{24} + \ldots \empe  
 &  \bmpe 1 - \frac{x^2}{2} + \frac{x^3}{3}
- \frac{x^4}{6} + \frac{x^5}{15} + \ldots \empe  \\ 
\hline
Chiral  ($\nu=0$) &  \bmpe 1 + \frac{x^2}{2}\log x + \ldots \empe  
&  \bmpe 1 + \frac{x^4}{16}\log x + \ldots \empe  \\
                 ($\nu=2/\beta$) &  \bmpe 1 - \frac{x^2}{4} + \ldots \empe  
&  \bmpe 1 - \frac{x^4}{64}+ \ldots \empe  \\
\end{tabular}
\caption{Expansion of the gap probabilities for small radii $x=Nr^2\ll1$.}
\label{lowx}
\end{table}
Therefore we also give the expansion of the gap probabilities close to
the origin. 
Note that the first product begins at zero, unlike in \cite{GHS}, as we
give the gap probability here in Table \ref{lowx}, compared to the
conditional probability to have one eigenvalue at zero and one at radius $r$
there. The first 
two equations agree with \cite{Mehta} eqs. (15.1.19) and (15.2.18). 

The leading power of the spacing distribution $p(s)$ can be reproduced for
Ginibre 
$\beta=2$ as follows. Multiplying the first line in Table \ref{lowx} with 
$e^{x}$ to remove the $n=0$ contribution also removes the linear term $-x$,
and  
we obtain $1-\frac12 x^2+\ldots$.
The spacing is then obtained by first reinserting $x=Nr^2$ and 
then differentiating with respect to $r$: $p(s)\sim s^3$. 
The same cubic repulsion was found in
\cite{Oas} for other examples of rotationally invariant weights, as well as in
\cite{FKS98} for strong non-Hermiticity by expanding the gap to second order as
in eq. (\ref{Eexp}), and it is thus considered to be universal. 
Comparing this to the chiral ensemble with one exact zero eigenvalue
$\nu=1$, we find once more a cubic repulsion, adding a further 
ensemble to this universality class.

While for $\beta=4$ a comparison to Ginibre is not possible we can at least
compare the two chiral ensembles, finding that the repulsion is much stronger
for $\beta=4$ than for $\beta=2$: 
here placing {\it one} eigenvalue at the origin corresponds to
$\nu=\frac12$. This indicates that there exists a different universality class
$p(s)\sim s^7$ of spacing distributions for $\beta=4$ at the origin.

As a final observation we note that when comparing ensembles in Table
\ref{lowx} 
the first power in $x$ doubles when going from Ginibre to chiral, or from
$\beta=2$ to $\beta=4$. This is in contrast to the coefficient in the exponent 
going down by $\frac12$ for these
comparisons of ensembles in the large-$x$ asymptotics in Table \ref{Presults}.

\sect{Conclusions}\label{conc}

We have investigated the gap probabilities and distributions of individual
eigenvalues with respect to radial ordering in the complex plane in
non-Hermitian Random Matrix Theory (RMT). 

After setting up a general framework in terms of Fredholm determinants and
Pfaffians for general non-Gaussian RMT with unitary ($\beta=2$)
and symplectic ($\beta=4$) invariance 
we turned to maximal non-Hermiticity. For general weights with rotational
invariance we found that the product representation of the gap probabilities 
in terms of
Fredholm eigenvalues are related for $\beta=2$ and 4, and we gave explicit new
expressions  for two ensembles of Gaussian chiral RMT. 
This relation between the gap
probabilities is different from the one for Hermitian RMT. It would be very
interesting to extend our relation to intermediate Hermiticity, interpolating
between the two limiting cases. This may be compared to the Hermitian limit
of the spectral correlators.  For $\beta=4$ these are given in terms of a 
$2\times2$ matrix kernel containing a 
single pre-kernel of skew orthogonal polynomials in the complex 
plane which is much simpler than the three pre-kernels in the Hermitian limit
following from a Taylor expansion of the complex one. 

We then derived an asymptotic expansion for the gap probability at large
radii for the Gaussian Ginibre and chiral complex ensembles, both at $\beta=2$
and 4. In particular this included a detailed discussion how to get from
Ginibre to chiral and from $\beta=2$ to 4 in these ensembles. 
Our results 
are consistent with the known results for Ginibre at $\beta=2$ of Forrester,
for the other ensembles they were new. 
It would be very interesting to sharpen our strict upper and lower
bounds for the linear coefficients, and we have given numerical evidence
for 
their conjectured values. Expanding for small radii we found that the 
chiral complex $\beta=2$ ensemble also displays a cubic level repulsion at the
origin, 
in contrast to its $\beta=4$ counterpart.

While in this paper we have focused on the $\beta=2$ and 4 ensembles one could
try to generalise our results to the recently solved $\beta=1$ Ginibre and
chiral ensemble with real asymmetric matrix entries. 
Because these two ensembles have both real and complex eigenvalues, several
different gap probabilities can be defined and computed, and we leave these
open questions for future work.

\indent

\noindent
\underline{Acknowledgements}:
Financial support by an EPSRC doctoral training grant (M.J.P.) and
EPSRC first grant EP/D031613/1 (G.A. and L.S.), as well as the
European Community Network grant ENRAGE MRTN-CT-2004-005616 (G.A.)
is gratefully acknowledged.
We thank Peter Forrester for kindly providing a reference to a little known
paper of his after we had written up this work.

\begin{appendix}

\sect{Integrals over Bessel functions}
\label{Kintegrals}

In this appendix we derive several integrals over Bessel functions, including
the one given in eq. (\ref{Kintb2}).

\subsection{General case}

First we compute the matrix elements of a determinant yielding the gap
probability $E_0^{(2)}(r)$ in the general elliptic case of the chiral
ensembles. For that purpose a slightly different representation than eq.
(\ref{gap2step1}) is more convenient.
Instead of choosing orthonormal wave functions in eq. (\ref{gap2step1})
one can simply keep the original monomials in the Vandermonde
determinant, leading to
\be
E_0^{(2)}(r) = \frac{N!}{{\cal Z}^{(2)}_{ch}}
\det_{1\leq k,j\leq N}\Big[ \int_{{\mathbb C}\setminus {\cal C}_r}d^2z\,
w_\nu^{(2)}(z) z^{2(k-1)}z^{*\,2(j-1)}\Big]
\label{gap2mehta}
\ee
instead of eq. (\ref{pregap2}) (the same identity was shown in a different way
in \cite{Mehta} eq. (15.1.13), without squared arguments).
While this seems tailored to the rotationally invariant case at
maximal non-Hermiticity \cite{Mehta}
we can also use it here in the general case.
We have to compute the following integrals:
\bea
&&\int_r^\infty ds\, s\int_0^{2\pi}d\theta s^{2\nu+2}
K_\nu(as^2)\exp\Big[bs^2\cos(2\theta)\Big]
s^{2(k+j-2)}\exp[2i\theta(k+j-2)]\ =\nn\\
&=&\pi\int_{r^2}^\infty dt\, t^{k+j+\nu-1} K_\nu(at) I_{k+j-2}(bt)\\
&&\mbox{with}\ \ \ \ 
a=\frac{N(1+\mu^2)}{2\mu^2}\ \ \ , \ b=  \frac{N(1-\mu^2)}{2\mu^2}\ ,
\eea
and we have $a>b\geq0$ ensuring convergence.
While we were unable to
give a general result we can provide a recursive prescription, starting with
the 
known integral
\be
\int_{r^2}^\infty dt\, t K_0(at) I_{0}(bt)= \frac{1}{a^2-b^2}\Big[
br^2I_{1}(br^2) K_0(ar^2)+ar^2I_{0}(br^2) K_1(ar^2)\Big] \equiv f(a,b;r^2)
\ ,
\label{fdef}
\ee
for $\nu=0$ and $m\equiv k+j-2=0$. Any value of $m>0$ and $\nu>0$
can be obtained from this by
applying the following Bessel identities for $a,b\neq0$
\bea
\partial_bI_m(bt)-\frac{m}{b} I_m(bt) &=& tI_{m+1}(bt)\nn\\
-\partial_aK_\nu(bt)+\frac{\nu}{a} K_\nu(at) &=& tK_{\nu+1}(at)\ .
\label{Besselid}
\eea

We thus obtain as a final result for $m=k+j-2$
\bea
&&\int_{r^2}^\infty dt\, t^{m+\nu+1} K_\nu(at) I_{m}(bt)\ =\nn\\
&=&\Big(\partial_b-\frac{m}{b}\Big)\Big(\partial_b-\frac{m-1}{b}\Big)\cdots
\partial_b \Big(-\partial_a+\frac{\nu}{a}\Big)
\Big(-\partial_a+\frac{\nu-1}{a}\Big)\cdots(-\partial_a)f(a,b;r^2)\ .
\eea
All differentiations can be carried out recursively and are algebraic,
acting on the expression in the middle of eq. (\ref{fdef}).
For $b=0$ or $\mu=1$ we have rotational invariance and all integrals can be
computed in a closed form, see Subsection \ref{max} and this appendix below.

We end this part on the elliptic case by sketching how to proceed for
$\beta=4$. Keeping monic powers instead of skew orthogonal polynomials in eq.
(\ref{pregap4}) we arrive at
\be
E_0^{(4)}(r) = \frac{(2N)!}{{\cal Z}^{(4)}}
\Pf_{1\leq k,l\leq 2N}\left[
\int_{{\mathbb C}\setminus {\cal C}_r}d^2z(z^2-z^{*\,2})w_\nu^{(4)}(z)
\Big( z^{2k-2}z^{*\,2l-2}- z^{*\,2k-2}z^{2l-2}\Big)
\right]
\label{gap4Mehta}
\ee
After multiplying out we have to compute the same types of integrals as
already done for $\beta=2$, with the index $2\nu$ of the Bessel-$K$ function
and weight now increasing in steps
of two instead.


\subsection{Maximal non-Hermiticity}

In the second part of this appendix we compute the integral needed in
the rotationally invariant case $\mu=1\ (b=0)$ given in
eq. (\ref{Kintb2}). In this case the angular integration diagonalises the
matrix, see (\ref{gap2mehta}), and the Bessel-$I$ functions from
above become simple powers. We have to show that:
\bea
F_\nu(k,x) &\equiv& \int_0^x ds\,s^{2k+\nu+1} K_\nu(s)\label{Fknudef}\\
&=& 2^{2k+\nu}(k+\nu)!k!\left(
1-\frac{x^{2k+\nu+1}}{2^{2k+\nu}(k+\nu)!k!}K_{\nu+1}(x)\right.
\label{Fknu}\\
&&-\left.
x\sum_{l=0}^{k-2}\frac{1}{(l+\nu+2)!l!}\left(\frac{x}{2}\right)^{2l+\nu+2}
K_{\nu+1}(x)
-x\sum_{l=0}^{k-1} \frac{1}{(l+\nu+1)!l!}\left(\frac{x}{2}\right)^{2l+\nu+1}
K_{\nu+2}(x)\right),
\nn
\eea
where the sums $\sum_{l=0}^{-2}$, $\sum_{l=0}^{-1}$ are set to zero.
For $k=0$ this integral is standard, see e.g. eq. (6.561.8) in \cite{Grad}
\be
\int_0^x ds\,s^{\nu+1} K_\nu(s)\ =\ 2^\nu\nu!-x^{\nu+1}K_{\nu+1}(x)\ ,
\ee
and in this case the sums giving the incomplete Bessel-$I$ functions
in the second line are absent.
Using the identities
\be
\Big(s^{\nu+1}K_{\nu+1}(s)\Big)^\prime\ =\ -s^{\nu+1}K_{\nu}(s)
\ \ \mbox{and}\ \ \Big( s^{-\nu}K_\nu(s)\Big)^\prime\ =\ -s^{-\nu}K_{\nu+1}(s)
\ ,
\ee
we can show that the following recursion holds:
\bea
F_\nu(k+1,x) &=&
-\int_0^x ds\,s^{2k+2}\Big(s^{\nu+1}K_{\nu+1}(s)\Big)^\prime
\label{Frec}\\
&=& -x^{2k+2+\nu+1} K_{\nu+1}(x) -  2(k+1)\int_0^x ds\,s^{2k+2+2\nu}
\Big( s^{-\nu}K_\nu(s)\Big)^\prime\nn\\
&=& -x^{2k+3+\nu} K_{\nu+1}(x) -2(k+1)x^{2k+2+\nu} K_{\nu}(x)+
4(k+1)(k+1+\nu)F_\nu(k,x)\ . \nn
\eea
It is easy to verify that the explicit expression eq. (\ref{Fknu})
satisfies this relation
\bea
F_\nu(k+1,x)
&=& 2^{2k+2+\nu}(k+1+\nu)!(k+1)!\left(
1-\frac{x^{2k+2+\nu+1}}{2^{2k+2+\nu}(k+1+\nu)!(k+1)!}K_{\nu+1}(x)\right.\nn\\
&&-
x\sum_{l=0}^{k-2}\frac{1}{(l+\nu+2)!l!}\left(\frac{x}{2}\right)^{2l+\nu+2}
K_{\nu+1}(x)
-x\frac{\left(\frac{x}{2}\right)^{2k-2+\nu+2}}{(k-1+\nu+2)!(k-1)!}K_{\nu+1}(x)
\nn\\
&&\left.
-x\sum_{l=0}^{k-1} \frac{1}{(l+\nu+1)!l!}\left(\frac{x}{2}\right)^{2l+\nu+1}
K_{\nu+2}(x)
-x \frac{1}{(k+\nu+1)!k!}\left(\frac{x}{2}\right)^{2k+\nu+1}\!K_{\nu+2}(x)
\right).
\nn
\eea
In a last step we have to apply the Bessel-$K$ identity
\be
2(\nu+1)K_{\nu+1}(x)-xK_{\nu+2}(x)\ =\ -xK_{\nu}(x)\ ,
\ee
to arrive at 
\bea
F_\nu(k+1,x)
&=& 4(k+1+\nu)(k+1)\Big(F_\nu(k,x) +x^{2k+\nu+1}K_{\nu+1}(x)\Big)
-x^{2k+3+\nu}K_{\nu+1}(x)\nn\\
&&-4(k+1)kx^{2k+\nu+1}K_{\nu+1}(x)-2(k+1)x^{2k+\nu+1}K_{\nu+2}(x)
\eea
which finishes our proof by induction.
As before the same integrals computed here apply to $\beta=4$.


\sect{A lower bound for $g_n^{Gin}(x)$ when $n > x$} \label{g_n_appendix}

We derive a lower bound for $g_n^{Gin}(x)$, valid when $n > x$. In order to do
this, we first need to determine an \textit{upper} bound for the individual
terms in $g_n^{Gin}(x)$, denoted $t_k(x)$, for $k > n > x$.  We will actually
do this for $k \ge x$. 


\subsection{An upper bound for the reciprocal of a factorial}

We have that \cite{Abramowitz}
\be
n! > \sqrt{2\pi n} \left( \frac{n}{e} \right)^n
\ee
and thus
\be
\label{inv_factorial_bound}
\frac{1}{n!} < \frac{1}{\sqrt{2\pi n}} \left( \frac{e}{n} \right)^n.
\ee


\subsection{An upper bound for $t_k(x)$} \label{T_k_appendix}

We define

\be
\label{T_k_definition}
t_k(x) \equiv \frac{e^{-x}x^k}{k!}
\ee
and wish to determine an upper bound for this when $k \geq x$. It proves
useful in  what follows to introduce a scaled variable 
\be
\label{p_definition}
p \equiv \frac{k-x}{\sqrt{x}} \geq 0.
\ee
We use eq. (\ref{inv_factorial_bound}), and write $k$ in terms of $p$ as
follows: 
\bea
t_k(x)  &<& \frac{e^{-x}x^k}{\sqrt{2\pi k}} \left( \frac{e}{k} \right)^k
\ =\ \frac{1}{\sqrt{2\pi x}} \exp \left[ p\sqrt{x} - \left(x +
p\sqrt{x} + \frac{1}{2}\right)\log\left( 1 + \frac{p}{\sqrt{x}} \right)
\right]    \nonumber \\ 
        &\equiv& \frac{1}{\sqrt{2\pi x}} \exp [ E(p,x) ]
\eea
Now, keeping $x$ fixed, we consider three different regimes for $k$
(or equivalently for $p$). 

First, for $x \leq k < \frac{3x}{2}$ (equivalently $0 \leq p <
\frac{\sqrt{x}}{2}$), we can expand the logarithm as an absolutely convergent
series, and collect powers of $x^{-1/2}$: 
\bea
E(p,x) &=& p\sqrt{x} - \left(x + p\sqrt{x} + \frac{1}{2}\right) \left\{
\frac{p}{\sqrt{x}} - \frac{p^2}{2x} + \frac{p^3}{3x^{3/2}} - \frac{p^4}{4x^2}
+ \ldots  \right\}   \nonumber \\ 
       &=& -\frac{p^2}{2} + \sum_{i=1}^{\infty} \frac{a_i(p)}{x^{i/2}}
\eea
where the coefficients $a_i(p)$ are polynomials in $p$:
\be
a_i(p) \equiv (-)^{i+1} p^i \left[ \frac{p^2}{(i+1)(i+2)} - \frac{1}{2i}
  \right]. 
\ee
Hence
\bea
t_k(x)  &<& \frac{1}{\sqrt{2\pi x}} \exp \left\{ -\frac{p^2}{2} +
\sum_{i=1}^{\infty} \frac{a_i(p)}{x^{i/2}} \right\} 
\ =\ \frac{1}{\sqrt{2\pi x}} e^{-p^2/2} \left\{ 1 + \sum_{i=1}^{\infty}
\frac{b_i(p)}{x^{i/2}} \right\} 
\ \equiv\ \overline{t}^{(1)}(k,x)\ \ 
\eea
where the $b_i(p)$ are also polynomials in $p$ which we do not give explicitly.

Second, for $\frac{3x}{2} \leq k < e^2 x$ (equivalently $\frac{\sqrt{x}}{2}
\leq p < (e^2-1)\sqrt{x}$), we use the fact that $t_k(x)$ is monotonic
decreasing in $k$ (for $k \geq x$ and $x$ fixed), to give 
\bea
t_k(x) &\leq& t_{\frac{3x}{2}}(x)  
\ =\ \frac{1}{\sqrt{2\pi x}} \sqrt{\frac{2}{3}} \exp \left\{ \left(
        \frac{1}{2} - \frac{3}{2} \log \frac{3}{2} \right) x \right\}
\ \equiv\ \overline{t}^{(2)}(x)\ \ 
\eea
which is (clearly) $p$-independent. The coefficient of $x$ in the exponent is
negative. 

Finally, for $k \geq e^2 x$ (equivalently $p \geq (e^2-1)\sqrt{x}$), we have
\be
\log\left( 1 + \frac{p}{\sqrt{x}} \right) \geq 2
\ee
and hence
\bea
t_k(x) &<&  \frac{e^{-2x-1}}{\sqrt{2\pi x}} e^{-\sqrt{x}\, p}   
\ \equiv\ \overline{t}^{(3)}(k,x)
\eea
in which we emphasise the $p$-dependence, since we will be integrating over
$p$. 


\subsection{A lower bound for $g_n^{Gin}(x)$}

Using the previous definitions of $g_n(x)$ and $t_k(x)$
(eqs. (\ref{g_n_definition}) and (\ref{T_k_definition}) respectively), we have
(for $n > x$) 
\bea
g_n^{Gin}(x) &\equiv&  \sum_{k=0}^n t_k(x)  
\ =\ 1 - \sum_{k=n+1}^{\infty} t_k(x)  \nonumber \\
       &>& \left\{\begin{array}{ll}
           1 - \int_n^{3x/2} \overline{t}^{(1)}(k,x) dk - \int_{3x/2}^{e^2x}
       \overline{t}^{(2)}(x) dk - \int_{e^2x}^{\infty} \overline{t}^{(3)}(k,x)
       dk  & (x<n<3x/2)  \\ 
           1 - \int_{n}^{e^2x} \overline{t}^{(2)}(x) dk - \int_{e^2x}^{\infty}
       \overline{t}^{(3)}(k,x) dk  & (3x/2 \le n < e^2x)\ \ \ \   \\ 
           1 - \int_{n}^{\infty} \overline{t}^{(3)}(k,x) dk  & (e^2x \le n)
\end{array}\right.  \\
       &>& 1 - \int_n^{\infty} \overline{t}^{(1)}(k,x) dk - \int_{3x/2}^{e^2x}
\overline{t}^{(2)}(x) dk - \int_{e^2x}^{\infty} \overline{t}^{(3)}(k,x) dk
\nonumber \\ 
\eea
where the $\overline{t}^{(i)}(k,x)$ 
are the upper bounds for $t_k(x)$ derived in
Appendix \ref{T_k_appendix}. We have also used the fact $t_k(x)$ is monotonic
decreasing as a function of $k$. 

In the same way that we introduced $p$ as a scaled proxy for $k$
(eq. (\ref{p_definition})), we similarly introduce $m$ as a scaled proxy for
$n$: 
\be
m \equiv \frac{n-x}{\sqrt{x}}
\ee
The first integral can then be written
\bea
\int_n^{\infty} \overline{t}_1(k,x) dk &=& \frac{1}{\sqrt{2\pi}}
\int_m^{\infty} e^{-p^2/2} \left\{ 1 + \sum_{i=1}^{\infty}
\frac{b_i(p)}{x^{i/2}} \right\} dp   \nonumber \\ 
   &=&  1 - \Phi(m) + \sum_{i=1}^{\infty} \frac{c_i(m)}{x^{i/2}}
\eea
where $\Phi(m)$ is the cumulative normal function, and the $c_i(m)$ have no
$x$-dependence. 

The second and third integrals have the forms $\alpha x^{1/2}e^{-\beta x}$ and
$\gamma x^{-1/2} e^{-\delta x}$ respectively, where $\alpha$, $\beta$,
$\gamma$ and $\delta$ are constants: 
\bea
\alpha &=& \frac{e^2-\frac{3}{2}}{\sqrt{3\pi}}, \nonumber \\
\beta  &=& \frac{3}{2}\log \left( \frac{3}{2} \right) - \frac{1}{2} > 0,
\nonumber \\ 
\gamma &=& \frac{1}{e\sqrt{2\pi}}, \nonumber \\
\delta &=& 1 + e^2 > 0. \nonumber \\
\eea
The important thing to note is that these two integrals are exponentially
small when compared with the first integral. 

Hence, we can write (as an exact result)
\be
g_n^{Gin}(x) \equiv g_{x+m\sqrt{x}}^{Gin}(x) \ >\ \Phi(m) - \sum_{i=1}^{\infty}
\frac{c_i(m)}{x^{i/2}} - \alpha \sqrt{x}e^{-\beta x} - \frac{\gamma e^{-\delta
    x}}{ \sqrt{x}}. 
\ee
This is true for any $m>0$ (equivalently, for any $n>x$). However, we see that
the right-hand side has a well-defined asymptotic limit as $x \rightarrow
\infty$ \textit{only for fixed $m$} (i.e. and not for fixed $n$). 


\end{appendix}

\newpage


\begin{thebibliography}{42}

\bibitem{Gin} J. Ginibre, J. Math. Phys. {\bf 6} (1965) 440.

\bibitem{FS} Y.V. Fyodorov and H.-J. Sommers,
{ J. Phys. {\bf A}: Math. Gen.} {\bf 36} (2003) 3303 [nlin.CD/0207051].

\bibitem{beta1list}
G. Akemann and E. Kanzieper, 
J. Stat. Phys. {\bf 129} (2007) 1159-1231 [math-ph/0703019];
H.-J. Sommers,   J. Phys. {\bf A40}, F671  (2007)
[arXiv:0706.1671];
P.J. Forrester and T. Nagao, Phys. Rev. Lett. {\bf 99}
  050603 (2007) [arXiv:0706.2020 [cond-mat.stat-mech]];
A. Borodin and C.D. Sinclair, arXiv:0706.2670v2 [math-ph];
arXiv:0805.2986 [math-ph];
P.J. Forrester and T. Nagao, J. Phys.  {\bf A41},
375003 (2008) [arXiv:0806.0055 [math-ph]];
H.-J. Sommers and W. Wieczorek,  J. Phys.  {\bf A41},
405003 (2008) [arXiv:0806.2756 [cond-mat.stat-mech]];
G. Akemann, M.J. Phillips, and H.-J. Sommers,
J. Phys. {\bf A}:
Math. Theor. {\bf 42} (2009) 012001 [ arXiv:0810.1458 [math-ph]].

\bibitem{ABSW}
G. Akemann, J. Bloch, L. Shifrin and T. Wettig,
Phys. Rev. Lett. {\bf 100} (2008) 032002
[arXiv:0710.2865v2 [hep-lat]];  PoSLAT2007: 224,2007
[arXiv:0711.0629v1 [hep-lat]].

\bibitem{GHS} R. Grobe, F. Haake and H.-J. Sommers, Phys. Rev.
Lett. {\bf 61} (1988) 1899.

\bibitem{Mehta}
M.L. Mehta, {\it Random Matrices}, Academic Press, Third
Edition, London 2004.

\bibitem{James} J.C. Osborn,
Phys. Rev. Lett. {\bf 93} (2004) 222001 [hep-th/0403131].

\bibitem{A05}
G.~Akemann,
  Nucl. Phys. {\bf B730} (2005) 253
  [hep-th/0507156].

\bibitem{A07mu} G. Akemann,
Int. J. Mod. Phys. {\bf A22} (2007) 1077 [hep-th/0701175]

\bibitem{MPW}
  H.~Markum, R.~Pullirsch and T.~Wettig,
  Phys.\ Rev.\ Lett.\  {\bf 83} (1999) 484
  [hep-lat/9906020].

\bibitem{dCM} J. des Cloizeaux and M.L. Mehta, J. Math. Phys. {\bf 14} (1973)
1648.

\bibitem{Igor} I. Krasovsky, Int. Math. Res. Not. {\bf 2004} (2004) 1249;
T. Ehrhardt, Comm. Math. Phys. {\bf 262} (2006) 317.

\bibitem{PF92} P. Forrester, Phys. Lett. {\bf A169} (1992)
21.

\bibitem{Sommers88} H.-J. Sommers, A. Crisanti, H. Sompolinsky and Y. Stein,
  Phys. Rev. Lett {\bf 60} (1988) 1895.

\bibitem{FKS} Y.V. Fyodorov, B.A. Khoruzhenko and H.-J. Sommers,
Phys. Lett. {\bf A226} (1997) 46  [cond-mat/9606173];
Phys. Rev. Lett. {\bf 79} (1997) 557  [cond-mat/9703152].

\bibitem{FKS98} Y.V. Fyodorov, B.A. Khoruzhenko and H.-J. Sommers,
Ann. Inst. Henri Poincar\'e
{\bf 68} (1998) 449 [chao-dyn/9802025].

\bibitem{EKb4} E. Kanzieper,
J. Phys. {\bf A}: Math. Gen. {\bf 35} (2002) 6631 [cond-mat/0109287].

\bibitem{AOSV}
G. Akemann, J.C. Osborn, K. Splittorff, and J.J.M. Verbaarschot,
Nucl. Phys. {\bf B712} (2005) 287
[hep-th/0411030].

\bibitem{ADp}
G.~Akemann and P.~H.~Damgaard,
  Phys.\ Lett.\  B {\bf 583} (2004) 199
  [hep-th/0311171].

\bibitem{Grad} I.S. Gradshteyn and I.M. Ryzhik, {\it Table of Integrals,
Series and Products}, 6th Edition, Academic
  Press, London (2000).

\bibitem{A03} G. Akemann,
J. Phys. {\bf A}: Math. Gen. {\bf 36} (2003) 3363 [hep-th/0204246].

\bibitem{Oas} G. Oas, Phys. Rev. {\bf E55} (1997) 205
[cond-mat/9610073].

\bibitem{Abramowitz} M. Abramowitz and I.E. Stegun, {\it Handbook of
Mathematical Functions}, Dover Publications Inc., Dover (1965).

\end{thebibliography}
\end{document}